\documentclass[aps,prl,twocolumn,floatfix,superscriptaddress]{revtex4-2}

\usepackage{graphicx}
\usepackage[english]{babel}
\usepackage{amsmath}
\usepackage{amssymb}
\usepackage{soul,color}

\begin{document}

\title{Custodial chiral symmetry in a Su-Schrieffer-Heeger electrical circuit with memory}

\author{Massimiliano Di Ventra}
\affiliation{Department of Physics, University of California, San Diego, La Jolla, CA 92093,
USA}
\author{Yuriy V. Pershin}
\affiliation{Department of Physics and Astronomy, University of South Carolina, Columbia,
South Carolina 29208, USA}
\author{Chih-Chun Chien}
\affiliation{Department of physics, University of California, Merced, CA 95343,
USA}
\email{cchien5@ucmerced.edu}

\begin{abstract}
Custodial symmetries are common in the Standard Model of 
particle physics. They arise when quantum corrections to a parameter are proportional to the parameter itself. Here, we show that a custodial symmetry of the chiral type is also present in a classical Su-Schrieffer-Heeger (SSH) electrical circuit with memory. In the absence of memory, the SSH circuit supports a symmetry-protected topological edge state. Memory induces nonlinearities that break chiral symmetry explicitly and spreads the state across the circuit. However, the resulting state is still protected against perturbations by the ensuing custodial chiral symmetry. These predictions can be verified 
experimentally and demonstrate the interplay between symmetry and memory.
\end{abstract}

\maketitle

A symmetry is said to be {\it custodial} if, despite being explicitly broken, it still protects physical quantities (e.g. the mass of particles) from 
large quantum corrections~\cite{Schwartz,Diaz01}. Symmetries of this type appear in the Standard Model of particle physics. They arise when quantum corrections to a parameter, as introduced by some symmetry-breaking term in the Lagrangian (e.g. a mass term), are proportional to the parameter itself. 

For instance, a custodial SU$(2)_V$ symmetry protects the mass relation between the electroweak $W$ and $Z$ gauge bosons from large quantum corrections; or a custodial chiral symmetry protects fermion 
masses from large radiative corrections. 

However, symmetry (like topology) is a concept that extends far beyond quantum systems. As such, it is natural to ask whether {\it custodial} symmetries can emerge in classical systems as well. In this Letter, we answer this question in the affirmative. In 
particular, we use the 1D Su-Schrieffer-Heeger (SSH) model with memory, as realized in electrical circuits with resistive memories~
\cite{PershinMemRev11}, as a prototypical example where this type of symmetry can be detected experimentally. The SSH model is a 
paradigmatic symmetry-protected topological insulator~\cite{Asboth2016,ShenTI}, namely it realizes a state of matter with a quantized topological indicator, known as the winding number, associated with a symmetry ({\it chiral} in the case of the SSH model) and a finite gap (in the thermodynamic limit). 

\begin{figure}[t]
    \centering
    \includegraphics[width=.9\columnwidth]{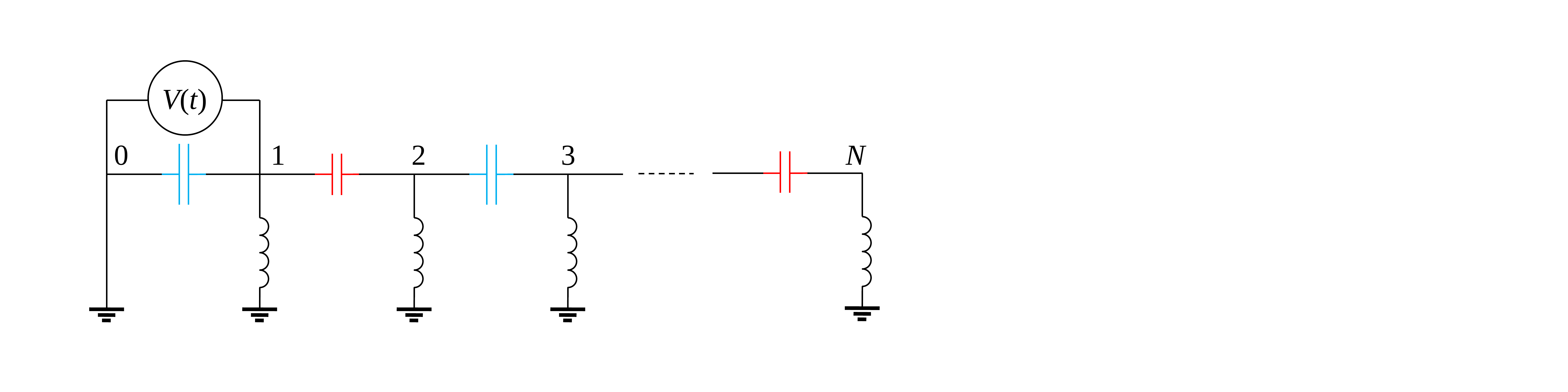} \\
    (a) \\
    \includegraphics[width=.9\columnwidth]{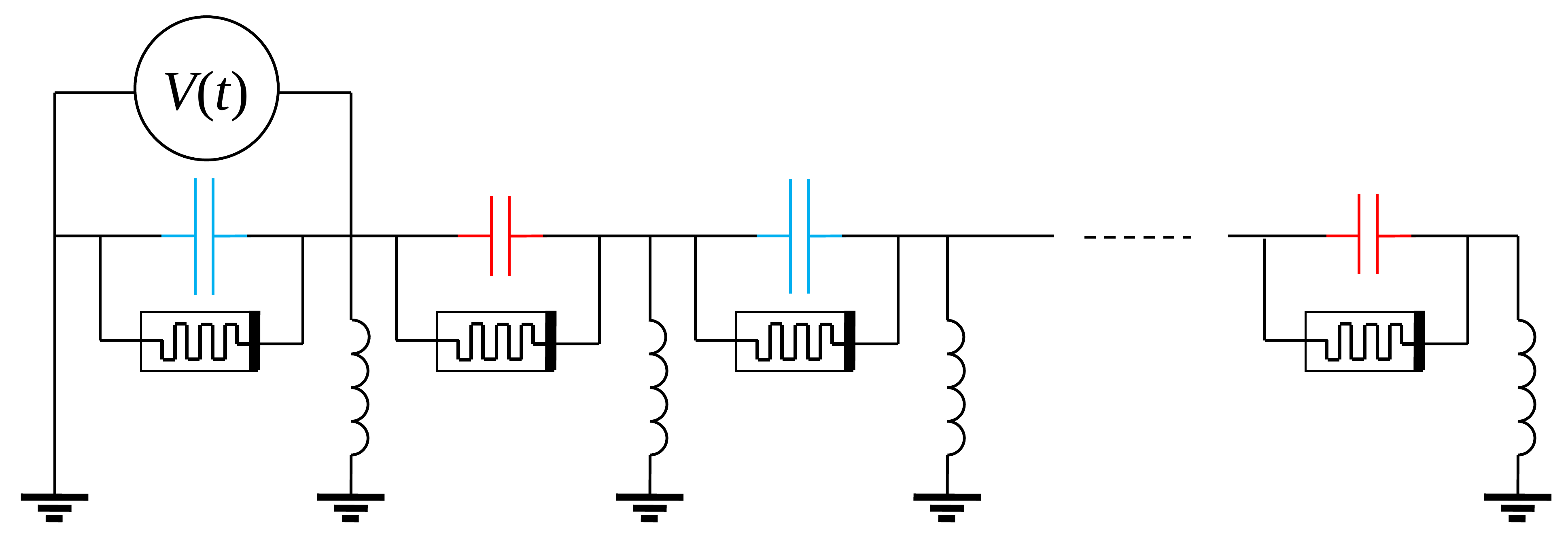} \\
    (b)
    \caption{Schematics of (a) standard and (b) memristive SSH circuit. The numbers in (a) represent the node number. In simulations, we used $C_1=0.22$~$\mu$F (blue (odd) capacitors), $C_2=0.1$~$\mu$F (red (even) capacitors), and $L=10$~$\mu$H (all inductors). \label{fig:1}}
\end{figure}

\begin{figure*}[tb]
    \centering
    \includegraphics[width=0.68\columnwidth]{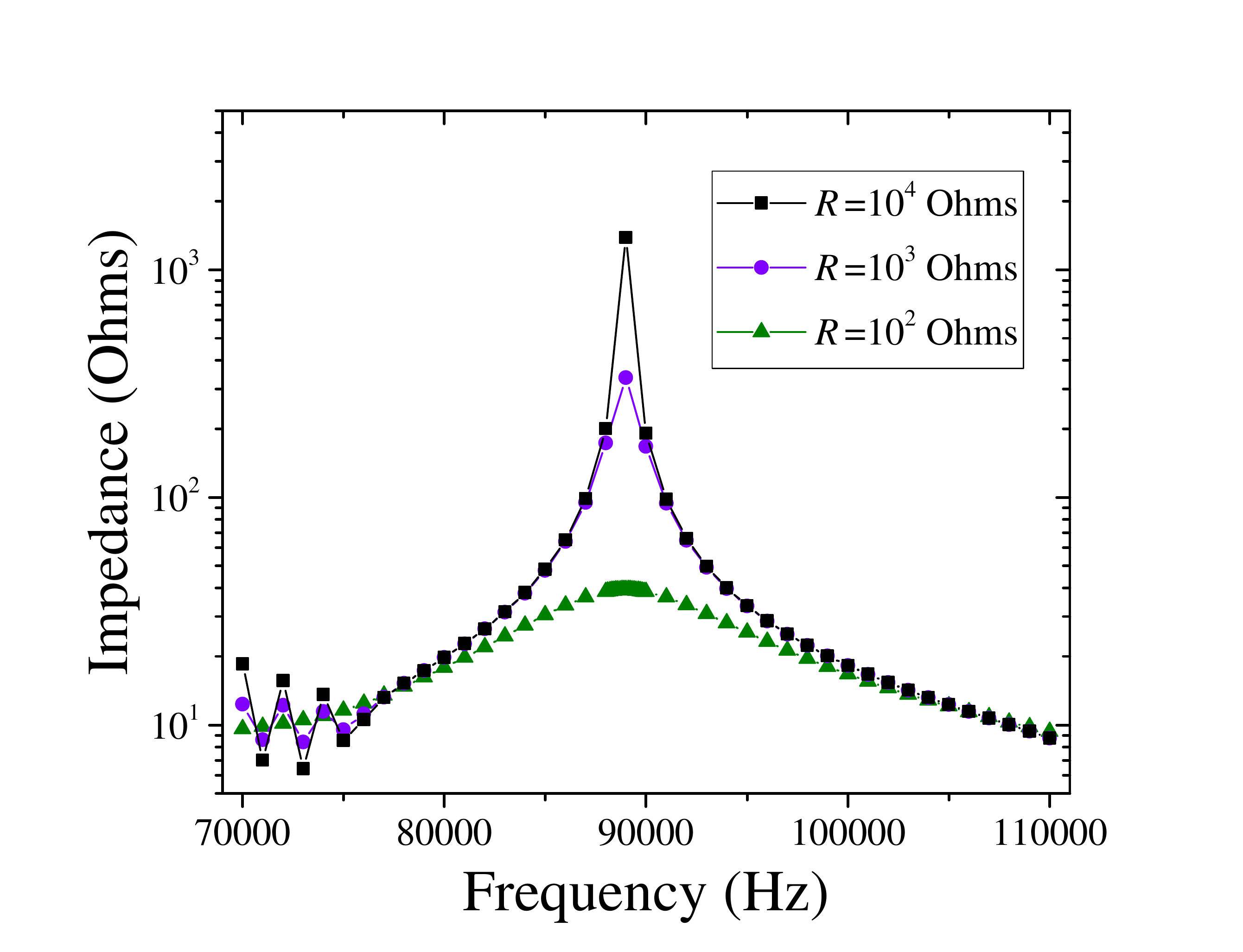} 
    \includegraphics[width=0.68\columnwidth]{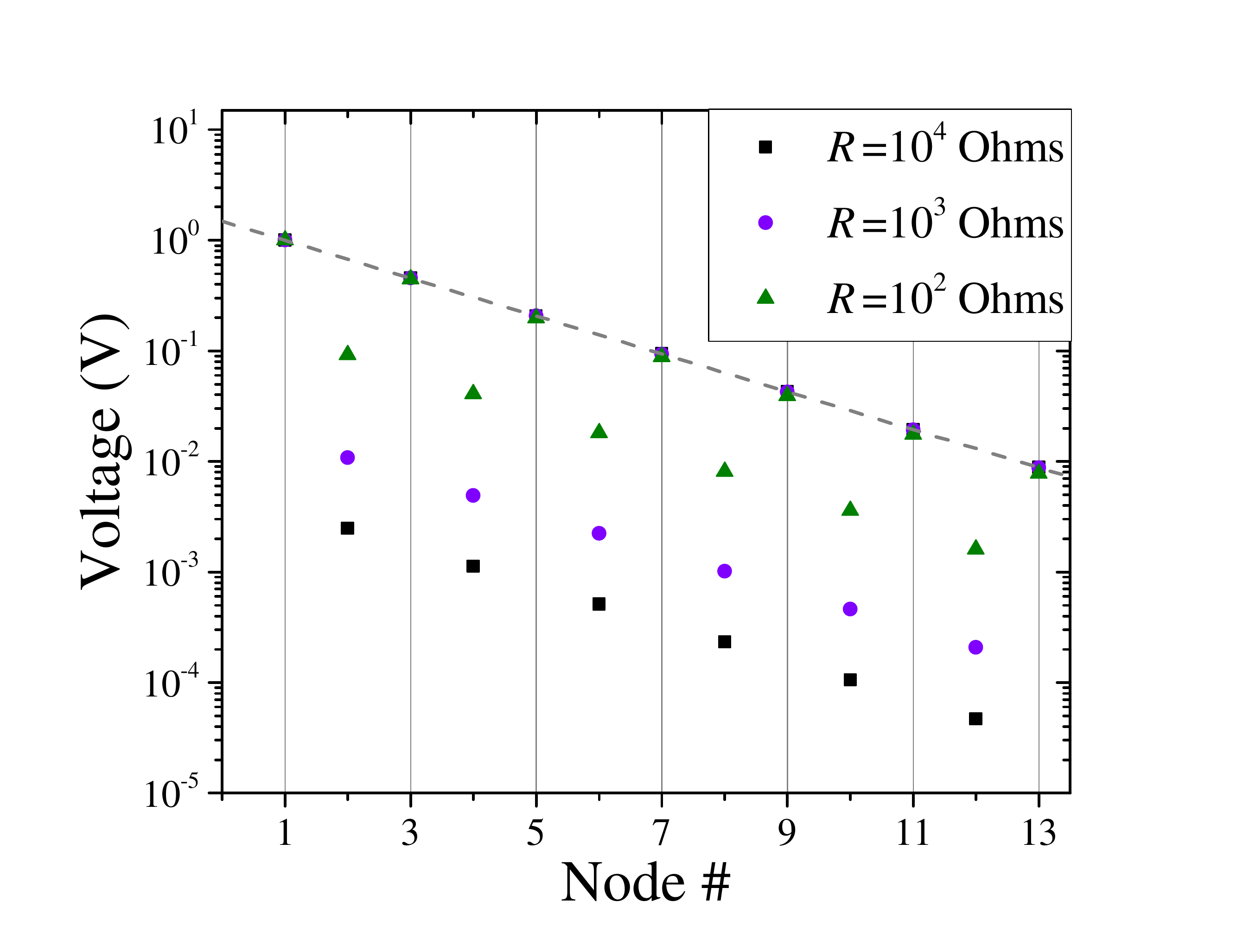} 
    \includegraphics[width=0.68\columnwidth]{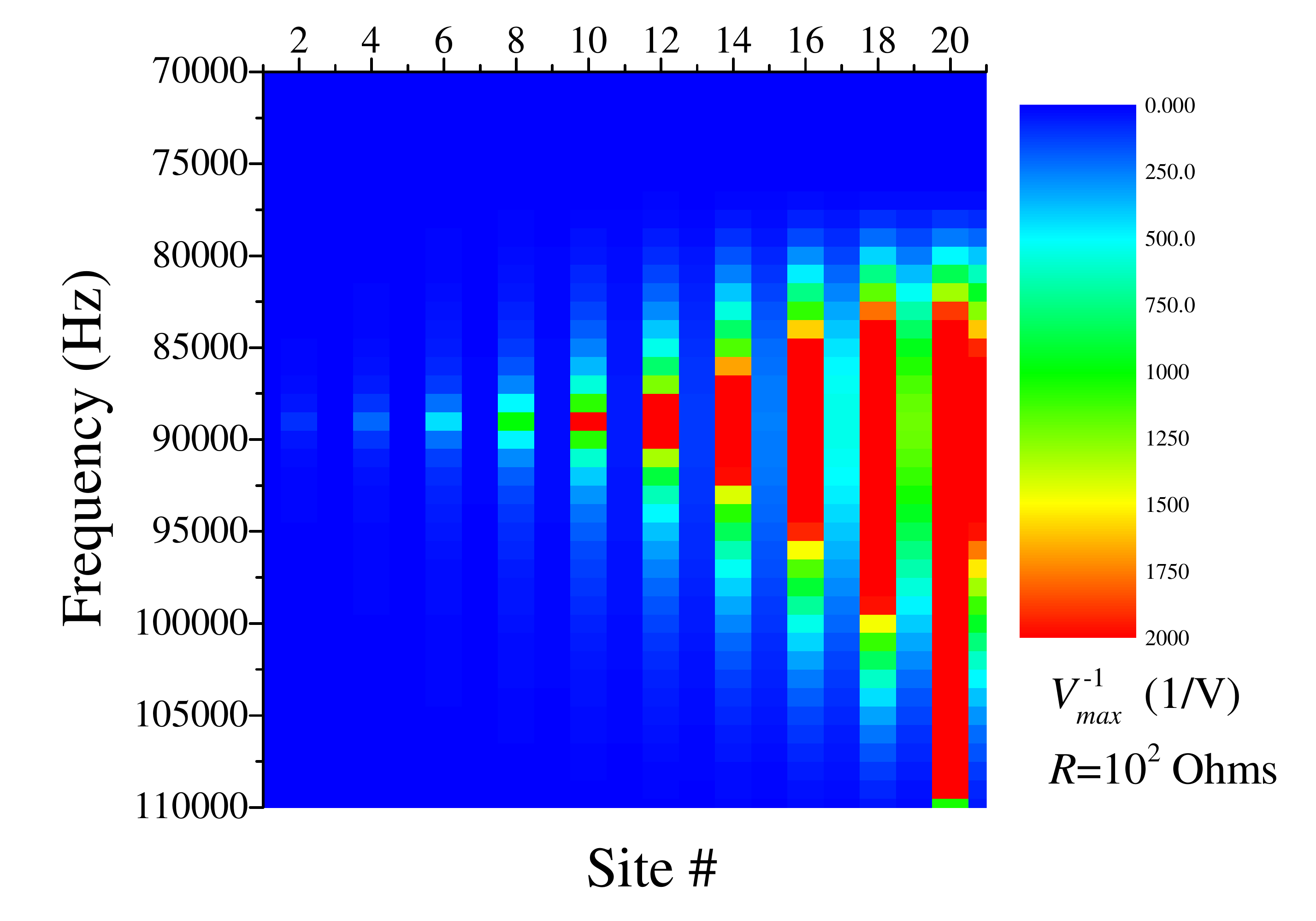} \\
    (a) \hspace{5cm} (b) \hspace{5cm} (c) \\
    \includegraphics[width=0.68\columnwidth]{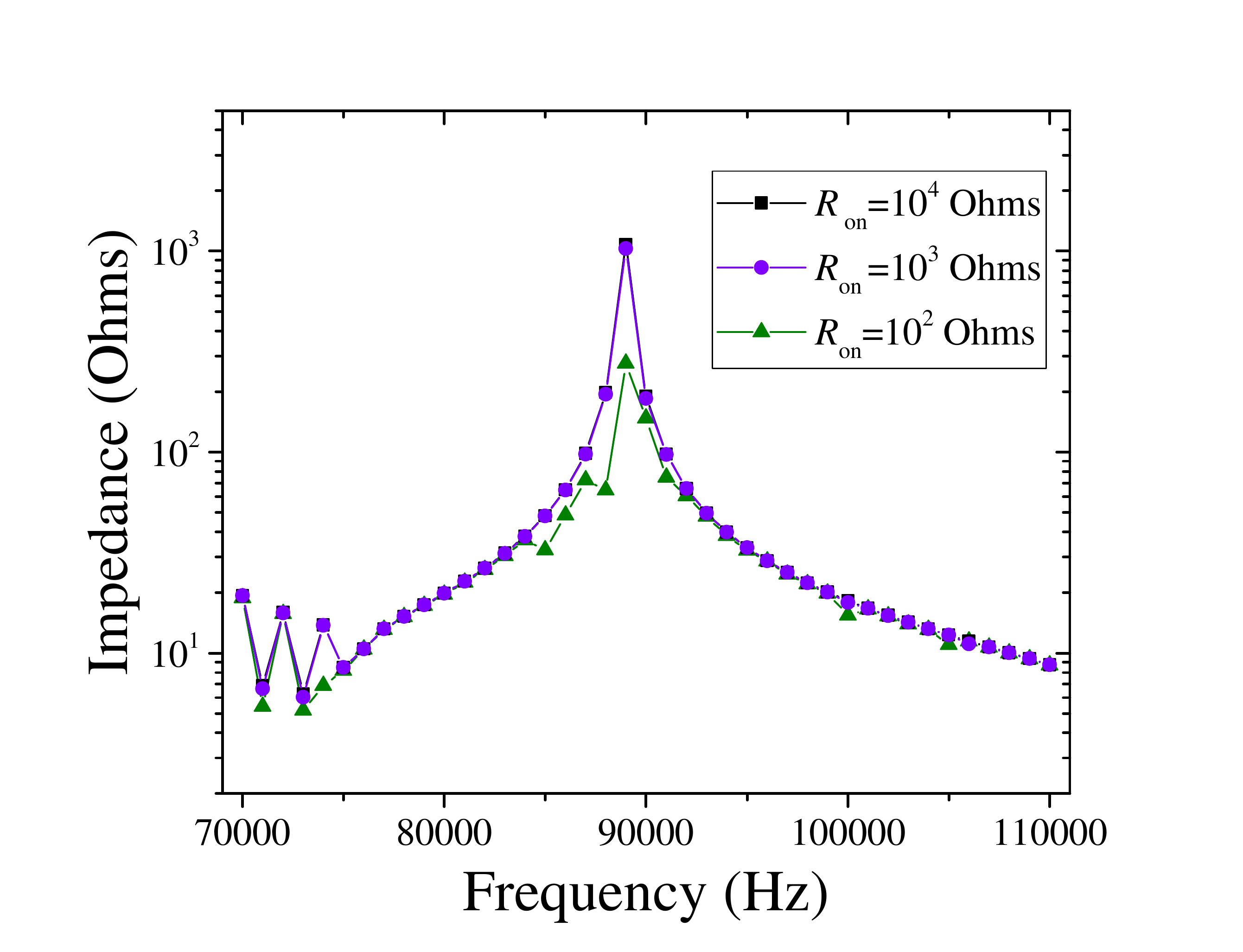} 
    \includegraphics[width=0.68\columnwidth]{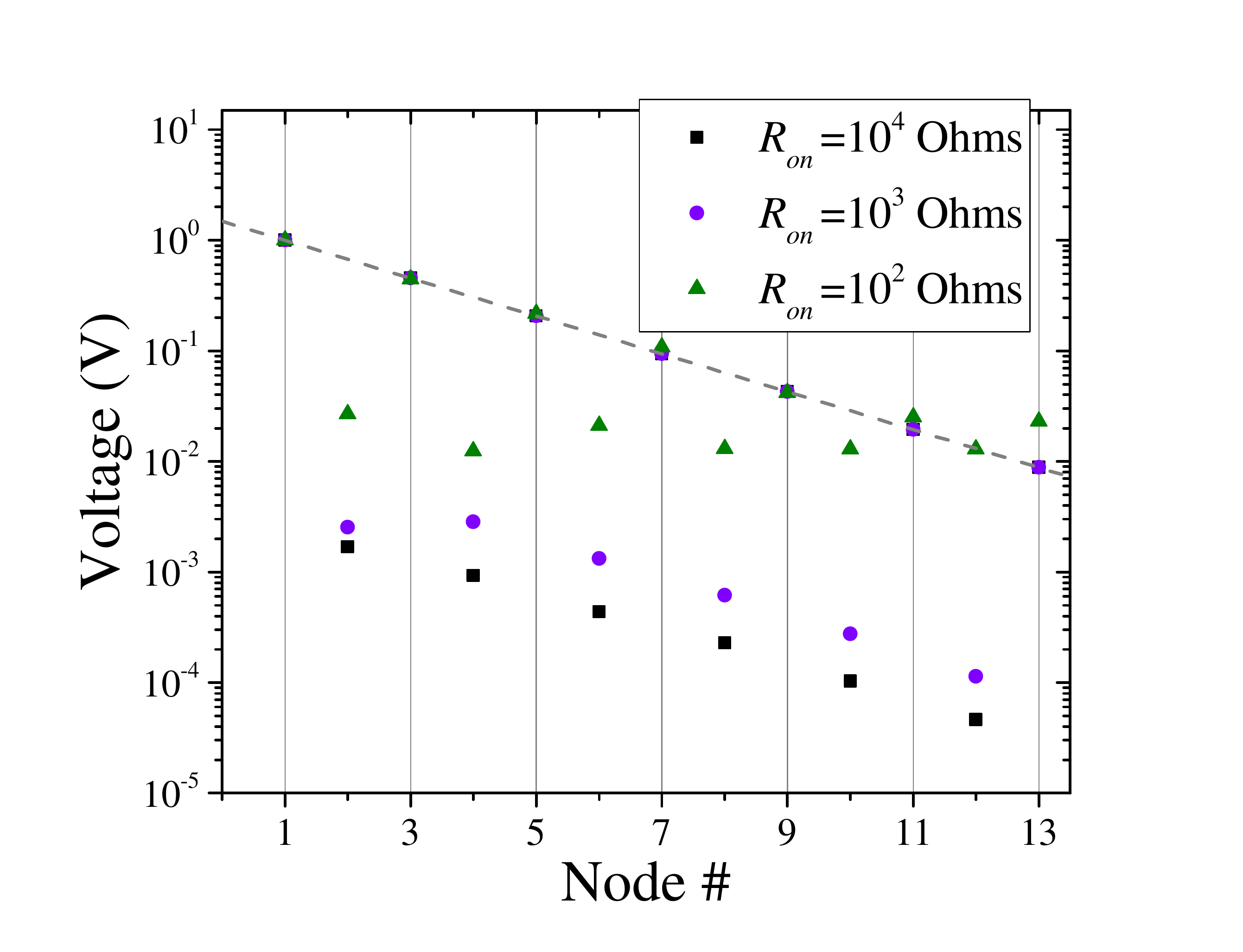}    \includegraphics[width=0.68\columnwidth]{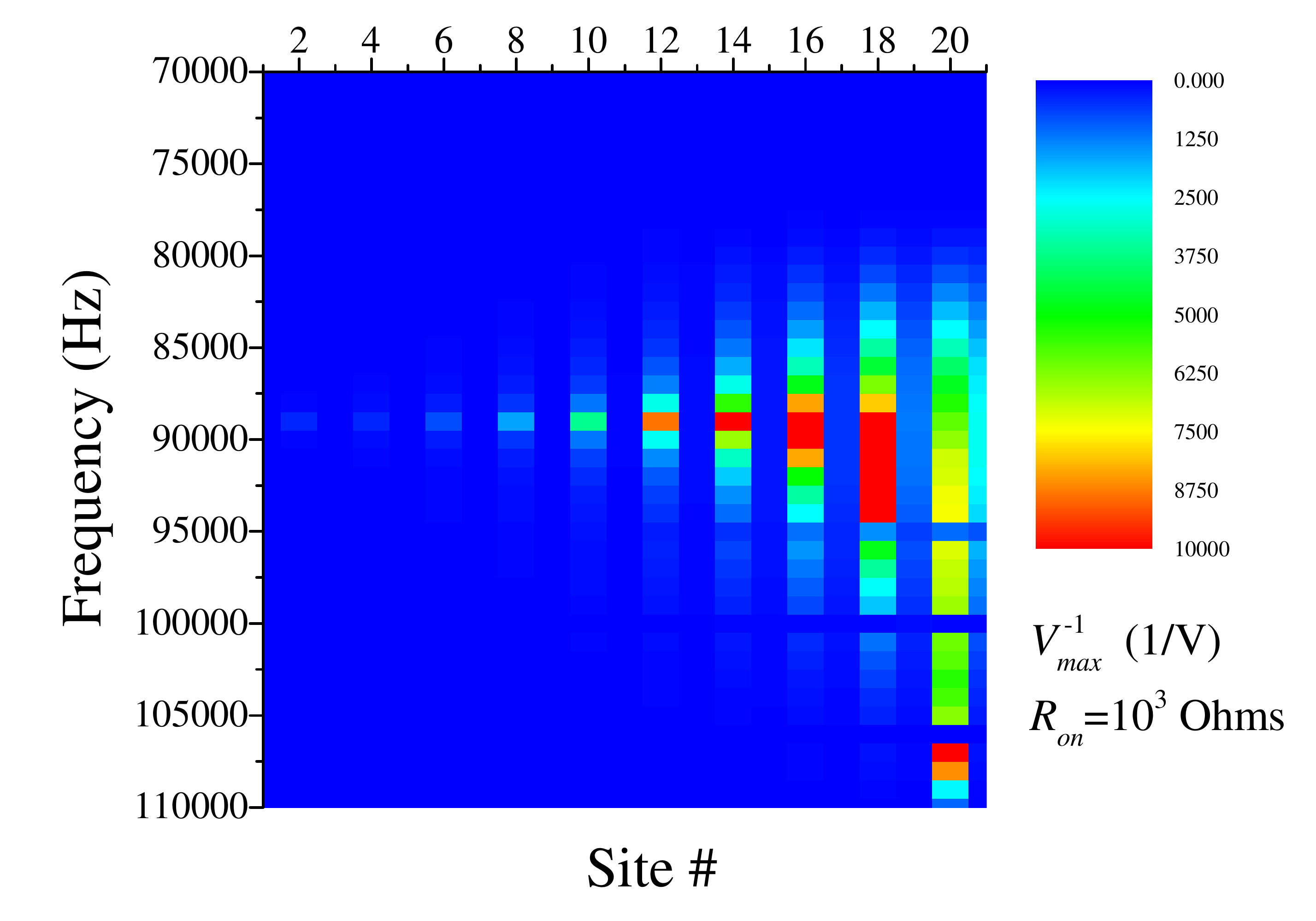} \\
    (d) \hspace{5cm} (e) \hspace{5cm} (f)
    \caption{Resistive/memristive SSH circuit simulations. (a) Impedance $Z_{10}$ of the circuit in Fig.~\ref{fig:1}(b) for different values of the resistance of standard resistors. (b) Maximum values of the node voltage at $\nu=89$~kHz. (c) Color map representation of $R=10^3$~Ohms calculation in (b). (d) Impedance of the circuit in Fig.~\ref{fig:1}(b) for different values of the low memristance state $R_{on}$. (e) Maximum values of the node voltage at $\nu=89$~kHz. We used $R_M(t=0)=10^4$~Ohms, $R_{off}=10^7$~Ohms, $\alpha=10^5$~(V$\cdot$s)$^{-1}$, $V_t=0.3$~V.
    (f) Color map representation of $R_{on}=10^3$~Ohms calculation in (e).
    \label{fig:2}}
\end{figure*}

In fact, there have been studies of electric circuits that simulate topological systems~\cite{Albert15,Ningyuan15,Imhof18,LeeTopolectric18,Garcia19,Pretko19,LiuRes19,Wang4D,Song20,YangPRR20,Hofmann20,Dong20,ZhangPRL21,Song21}. 
The simplest such circuit can be realized with an alternating 
series of capacitors, $C_1$ and $C_2$, and inductors, $L$ (Fig.~\ref{fig:1}). The ratio $C_2/C_1$ controls the existence of a symmetry-protected topological mid-gap state 
at one edge of the circuit. This state can be easily detected as a peak in the impedance as a function of frequency (Fig.~\ref{fig:2}a),
and is localized at the edge of the circuit, as a plot of the voltage drop at the nodes easily shows (Fig.~\ref{fig:2}b). This state is robust against local perturbations that do {\it not} break the chiral symmetry. We would then expect that if we introduced elements in the circuit that explicitly break such a symmetry, the edge state would disappear as the perturbation strength increases. 

Here, we introduce such elements in the form of experimentally-realizable resistors with memory (memristive elements) \cite{PershinMemRev11} in parallel with the capacitors (Fig.~\ref{fig:1}(b)). (The case of memristive elements in series with the capacitors is reported in the Supplementary Material.) Such elements introduce non-Hermiticity and strong nonlinearities, and break chiral symmetry explicitly. In fact, they delocalize the mid-gap state along the whole circuit. However, the resulting state is 
still robust against perturbations. We will show that the reason for this robustness is the reduction of the original chiral symmetry to a custodial status when memristive elements are added. 

{\it SSH circuit --} One way of emulating symmetry-protected topological systems via electric circuits~\cite{LeeTopolectric18} is based on the relation $\mathbf{I}=\mathbf{G}\mathbf{V}$ for idealized linear and memoryless elements. Here, $\mathbf{I}$ and $\mathbf{V}$ are the profiles of the current and voltage along the circuit, and $\mathbf{G}$ is the conductance matrix. 
By connecting alternating capacitors $C_1$ and $C_2$ in a circuit like the one shown in Fig.~\ref{fig:1}a, a classical analogue of the 1D SSH model~\cite{LeeTopolectric18}, can be realized. The SSH circuit with periodic boundary condition leads to the following conductance matrix in the Bloch form:
\begin{eqnarray}\label{eq:JSSH}
\mathbf{G}=i\omega[d_x(k)\sigma_x + d_y(k)\sigma_y + h_0 \mathbf{1}_2].
\end{eqnarray}
Here $d_x(k)=-[C_1+C_2\cos(k)]$, $d_y(k)=-C_2\sin(k)$, $h_0=[-\frac{1}{\omega^2  L}+C_1+C_2]$, $\mathbf{1}_2$ is the $2 \times 2$ identity matrix, and $\mathbf{\sigma}_{x,y,z}$ are the Pauli matrices. The momentum $k$ is within the first Brillouin zone. An inductor $L$ connected to the ground has been included in each cell. The conductance matrix~(\ref{eq:JSSH}) thus plays the role of the Hamiltonian.
In the ideal situation, namely in the absence of memory, dissipation, and/or other chiral symmetry-breaking elements, the eigenvalues of the conductance matrix resemble the energy spectrum of the quantum SSH model~\cite{LeeTopolectric18}. The two energy bands of the quantum model 
become, in the electrical circuit analogue, two continua of the conductance as a function of the driving frequency $\omega$.

As mentioned, the mid-gap state may emerge when the lattice has open boundary condition (see Fig.~\ref{fig:1}). In this case, a localized edge state emerges near the left edge if $C_2/C_1<1$, and can be detected experimentally by measuring, e.g., the two-point impedance,  $Z_{s0}=(V_s - V_0)/I$, between the source, $s$, and ground,  $0$, nodes of the circuit when the total current $I$ flows through it. The impedance then exhibits a peak with continua on both sides (Fig~\ref{fig:2}a).  
The localized state is associated with the winding number obtained from the equivalent system with periodic boundary condition via the bulk-boundary correspondence~\cite{Asboth2016}. The band (or bulk) states, meanwhile, contribute to clusters of small peaks on both sides and away from the localized-state peak (see the Supplementary Material). We will focus on the in-gap localized state hereinafter.

To more clearly see the emergence of the localized edge state, let us write explicitly the conductance matrix for the SSH circuit shown in Fig.~\ref{fig:1} with $N$ unit cells, each containing two capacitors and one inductor. Since the voltage in Fig.~\ref{fig:1} is applied to node $1$ while node $0$ is grounded, the conductance matrix starting from node 1 has the form $\mathbf{G}=A\mathbf{1}_{N}-i\omega \mathbf{B}$, where $A=i\omega[-(\omega^2 L)^{-1}+C_1+C_2]$ contributes to a uniform shift of the spectrum, and 
\begin{equation}\label{Bmatrix}
\mathbf{B}=\left(\begin{array}{ccccc}
0 & C_2 & 0 & 0 &\cdots \\
C_2 & 0 & C_1 & 0 &\cdots \\
0 & C_1 & 0 & C_2 & \cdots \\
\vdots & \vdots & \vdots & \vdots & \ddots
\end{array}
\right).
\end{equation}
By defining $\mathcal{C}=-C_2/C_1$, a voltage profile $V_{loc}=(V_1,V_2,\cdots)^T=(1, 0, \mathcal{C}, 0, \mathcal{C}^2, 0, \cdots)^{T}$ is an eigenstate of $\mathbf{G}$ at the resonant frequency $\tilde \omega = 1/\sqrt{L(C_1+C_2)}$. Here the superscript $T$ denotes the transpose. When $|\mathcal{C}|<1$, we then clearly see that the voltage profile shows an exponential decay towards the boundary opposite to the voltage source~\cite{IVnote}. Therefore, $V_{loc}$ is a localized state at the left edge
of the circuit.  This state can be seen in Fig~\ref{fig:2}b (see also Fig.~\ref{fig:220} in the Supplementary Material) in the configuration when the current is applied only to the first node. In addition, the diagonal part of the above matrix is uniform and can be separated out, indicating that the chiral (sublattice) symmetry is 
respected. To see this, let us define the following projection operators $P_1$ and $P_2$ into the odd and even sites, respectively:  $P_1=\textrm{diag}(1,0,1,0,\cdots)$ and $P_2=\textrm{diag}(0,1,0,1,\cdots)$. Then, $(P_1-P_2)\mathbf{B}(P_1-P_2)=-\mathbf{B}$ and the matrix $\mathbf{B}$ respects chiral symmetry.

{\it SSH model with memory --} We now show that this picture of band-topology/chiral symmetry changes dramatically in the presence of strong nonlinear effects in the form of, say, memristive elements in parallel with the capacitors (Fig.~\ref{fig:1}b)
To this end, let us employ a widely used model of memristive elements that reproduces the main features of experimentally-realized devices~\cite{PershinPRE09,PershinMemRev11}:
\begin{eqnarray}
R_M &=& R_{off}\left( 1-x \right)+R_{on}x , \label{memristive} \\
\frac{\textnormal{d}x}{\textnormal{d}t}&=&\alpha[\theta(V-V_t)(V-V_t)+\theta(-V-V_t)(V+V_t)].\;
\label{memristive1}
\end{eqnarray}
Here $R_M$ is the memory resistance, $R_{on}$ and $R_{off}$ are its limits, 
$x\in\left[0,1\right]$ is the internal state variable, $V_t$ is a positive threshold voltage, $\theta(v)$ is the Heaviside step function, and $\alpha$ is the switching rate. When combined in parallel with the capacitor and driven by $V=V_0 e^{i\omega t}$, we have $I=[i\omega C + (1/R_M)]V$ for a capacitor-resistor pair. Therefore, we can group the contributions from the resistors by defining $\tilde{C}_n=C_n+(i\omega R_{M,n})^{-1}$ for $n=1,2$, where $R_{M,n}$ is the memory resistance of the $n$-th memristive elements. This implies that  $I_n=i\omega\tilde{C}_n V_n$. Hence, one may replace $C_1,C_2$ in Eq.~\eqref{eq:JSSH} by $\tilde{C}_1, \tilde{C}_2$. 

Due to the memory effect from the memresistive elements, a full analysis of the system requires the (numerical) integration over time. However, to extract the main features observed in the simulations, we will use an approximate Fourier analysis (justified in the Supplementary Material). By generalizing $\mathbf{G}$ to the admittance matrix and assuming the form $\mathbf{I}=\mathbf{G}\mathbf{V}$ still holds for a frequency component, the first thing we notice is that the presence of the memristive elements leads to a {\it complex}-valued conductance matrix because the modified circuit dissipates energy. 
If we write again $\mathbf{G}=d_x\sigma_x +d_y\sigma_y+h_0 \mathbf{1}_2$, we then see that the trajectory $(d_x, d_y)$ now becomes a path in a {\it complex} space due to the presence of the memory elements, namely in a 4-dimensional (real) space, not 2-dimensional as before. While a loop in a 2-dimensional plane enclosing the origin cannot be smoothly transformed into a loop that does not enclose the origin, a loop in three or higher dimensions can smoothly deform around the origin without obstruction. 

Therefore, the winding number of the original (memoryless) SSH model is no longer faithful in characterizing the topology of the (non-Hermitian) memcircuit. Moreover, nonlinear effects come from the dependence of the memory resistance $R_M$ on the voltage (Eq.~(\ref{memristive})). While it is possible to characterize the topology of some non-Hermitian systems by using a biorthogonal basis~\cite{Ghatak19,Ashida21}, the presence of nonlinear effects in the memory circuit invalidates the construction of a basis for linear superposition~\cite{NonlinearNote}.
Therefore, the SSH circuit with memory elements defies the construction of conventional topological quantities. 

On the other hand, localized edge states in tight-binding models, known as Shockley-Tamm states~\cite{Shockley39,Tamm32}, may still arise due to symmetry and are not necessarily associated with the band topology. For a 1D lattice with nearest-neighbor interactions and alternating site strengths, the system has a {\it chiral} (or sublattice) symmetry. As explained previously, the symmetry can be observed by either constructing the projectors $P_A, P_B$ in real space or checking if the Bloch Hamiltonian anticommutes with an operator. 
For the SSH circuit with periodic boundary condition, $\sigma_z$ anticommutes with the diagonal part of the conduction matrix. Therefore, the chiral symmetry leads to pairs of the eigenstates. For the system shown in Fig.~\ref{fig:1}, a localized state may emerge with its energy pinned inside the bandgap in order to be consistent with the chiral symmetry. 
If we had regular resistors ($R_M=R=\textrm{constant}$) in parallel with the capacitors, we would introduce dissipation, rendering the system non-Hermitian, invalidating the winding number. However, the SSH circuit with regular resistors remains linear, and chiral symmetry still holds after the uniform diagonal part is factored out. Therefore, the edge state is still protected by the chiral symmetry~\cite{SSHEdgeNote}, and the impedance shows a broadened peak until the parallel resistance $R$ is small enough that the current will bypass the capacitors altogether (see Fig.~\ref{fig:2}a, where the peak decreases dramatically when we reduce $R$ from $10^4$ Ohms to $100$ Ohms).

{\it Custodial chiral symmetry --} If we now introduce memory into the resistors in parallel with the capacitors (Fig.~\ref{fig:1}b) two additional effects emerge. The nolinearity 
introduced by these elements breaks chiral symmetry explicitly: $\sigma_z \mathbf{B} \sigma_z\neq -\mathbf{B}$, where $\mathbf{B}$ is the admittance matrix after the uniform diagonal part is removed. The SSH edge state is
then no longer an eignestate of the conduction matrix. Nonetheless, we find the original SSH edge state is still present (as seen in the impedance curve of Fig.~\ref{fig:2}c) but {\it spreads} across the circuit 
(Fig.~\ref{fig:2}d). However, this effect emerges from the {\it diagonal} component of the conductance matrix. 
To see this, let us add alternating memristive elements $R_{M,1}$ and $R_{M,2}$ to the circuit~\cite{notepolarity}.
As discussed before, this leads to the effective capacitors with $\tilde{C}_{i}(x_j)=C_{i}(x_j)+[i\omega R_{M,i}(x_j)]^{-1}$, with $i=1,2$. Here $x_j$ labels the node location of the element with voltage $V(x_j)$ as shown in Fig.~\ref{fig:2}(d). 

To the lowest order in $R_{M,i}$, we may use $V_{loc}$ as the input voltage to get the profile of $R_{M,i}(x_j)$. The inhomogeneous $R_{M,i}(x_j)$ then leads to $\mathbf{G}=\tilde{A}(x_1)\mathbf{1}_{N}+\tilde{\mathbf{B}}$. Here $\tilde{A}(x_j)=i\omega[-(\omega^2 L(x_j)^{-1}+\tilde{C}_L(x_{j-1})+\tilde{C}_R(x_{j+1})]$, where $\tilde{C}_L$ and $\tilde{C}_R$ are the capacitor to the left and to the right of the node, and
\begin{equation}\label{Btilde}
\tilde{\mathbf{B}}=\left(\begin{array}{ccccc}
0 & \tilde{C}_2(x_2) & 0 & 0 &\cdots \\
\tilde{C}_2(x_1) & \alpha_1 & \tilde{C}_1(x_3) & 0 &\cdots \\
0 & \tilde{C}_1(x_2) & \alpha_2 & \tilde{C}_2(x_4) & \cdots \\
\vdots & \vdots & \vdots & \vdots & \ddots
\end{array}
\right).
\end{equation}
Here, $\alpha_j=\tilde {A}(x_{j+1})-\tilde{A}(x_1)$. 
The eigenstate of the conductance matrix is now
\begin{equation}\label{Vdis}
  V_{dis}=(1, 0, -\frac{\tilde{C}_{2}(x_1)}{\tilde{C}_{1}(x_3)}, \alpha_2\frac{\tilde{C}_2(x_1)}{\tilde{C}_1(x_3)\tilde{C}_2(x_4)},\cdots)^T,
\end{equation}
spreading over the whole circuit, consistent with the results of Fig.~\ref{fig:2}d.
By comparing the matrix $\tilde{\mathbf{B}}$ above with the matrix $\mathbf{B}$ of Eq.~(\ref{Bmatrix}), we see that the nonlinear terms violate the uniform diagonal of the linear case, thus breaking chiral symmetry. 
\begin{figure}[t!]
    \centering
    (a)\includegraphics[width=0.8\columnwidth]{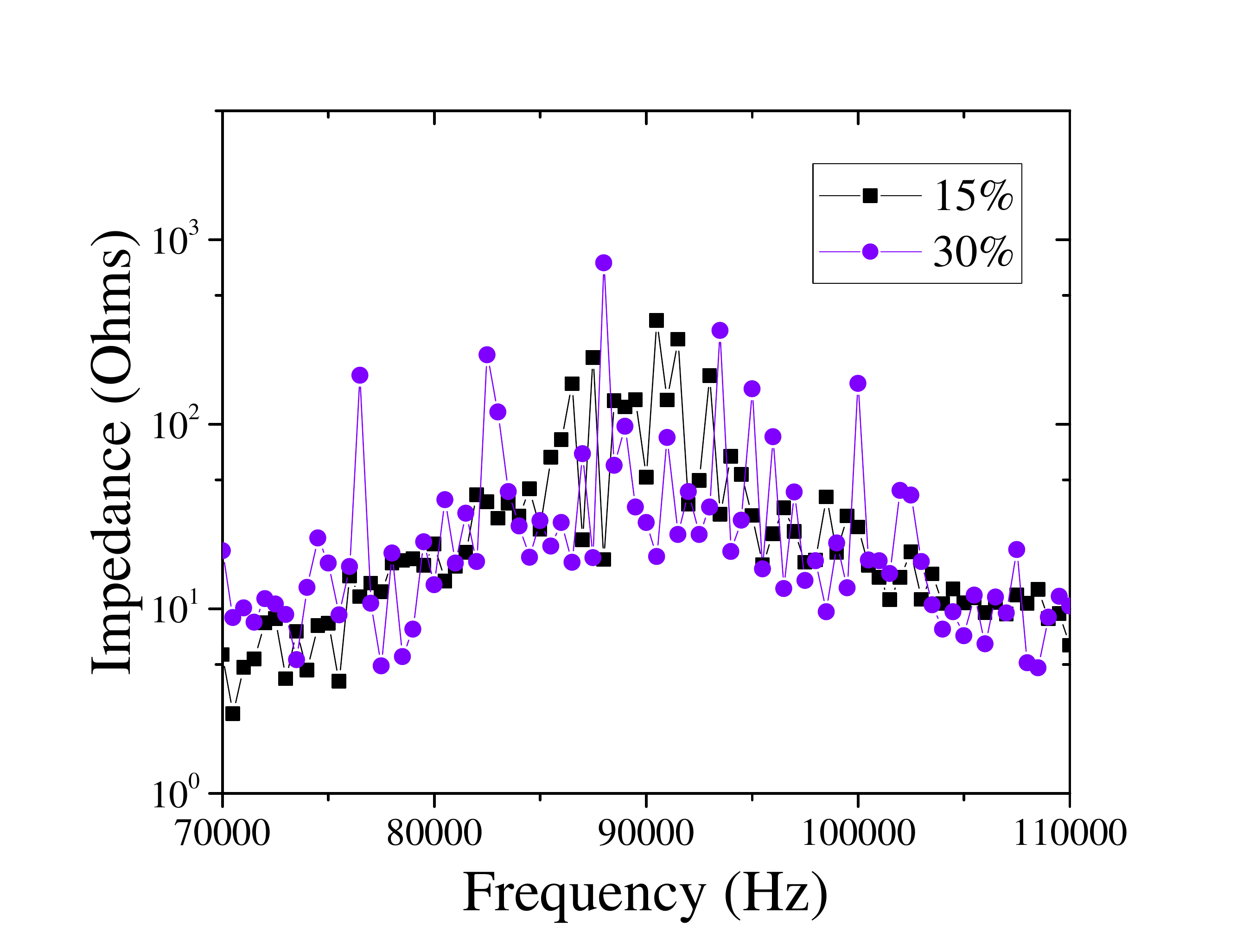} \;
    (b)\includegraphics[width=0.8\columnwidth]{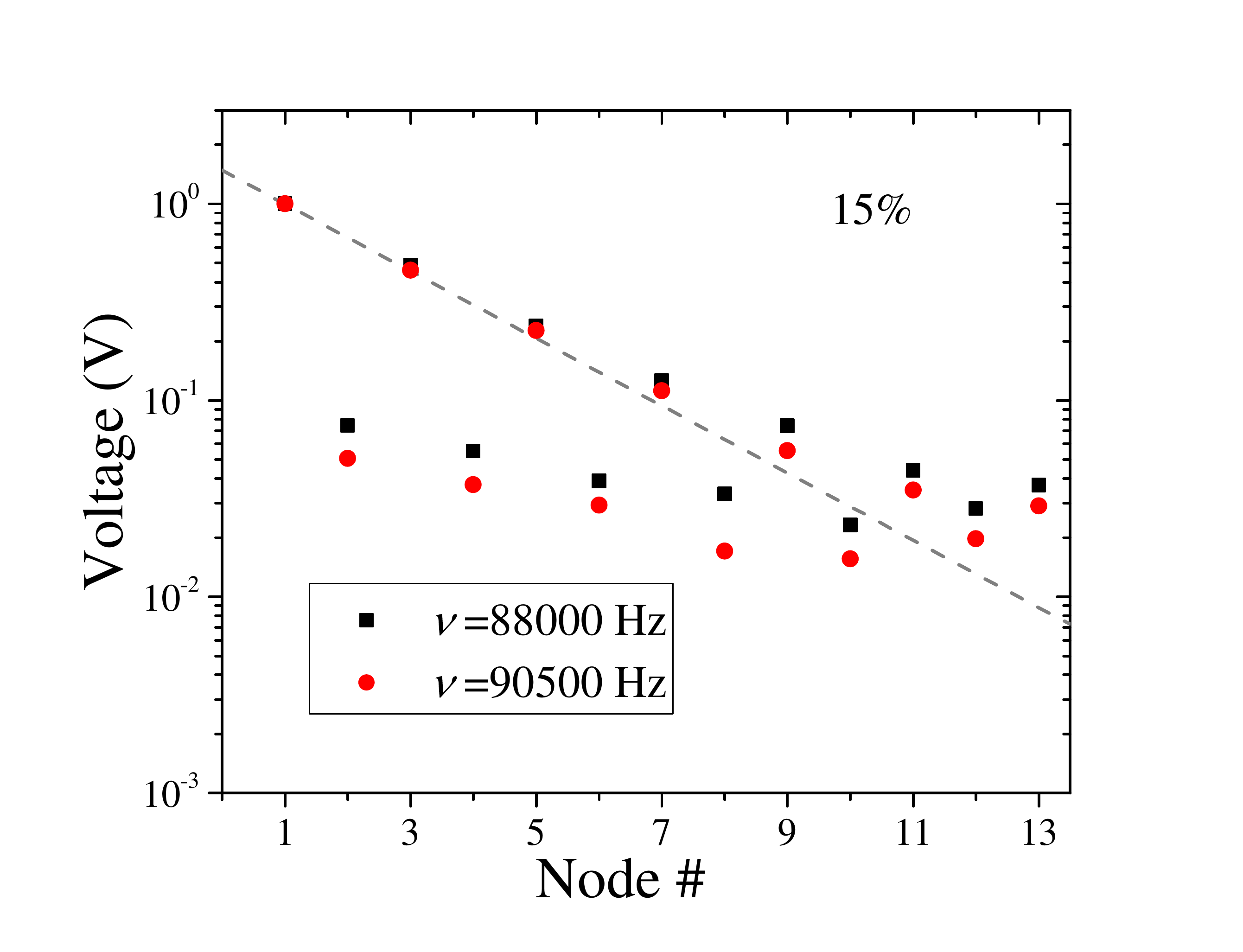} \\
    (c)\includegraphics[width=0.8\columnwidth]{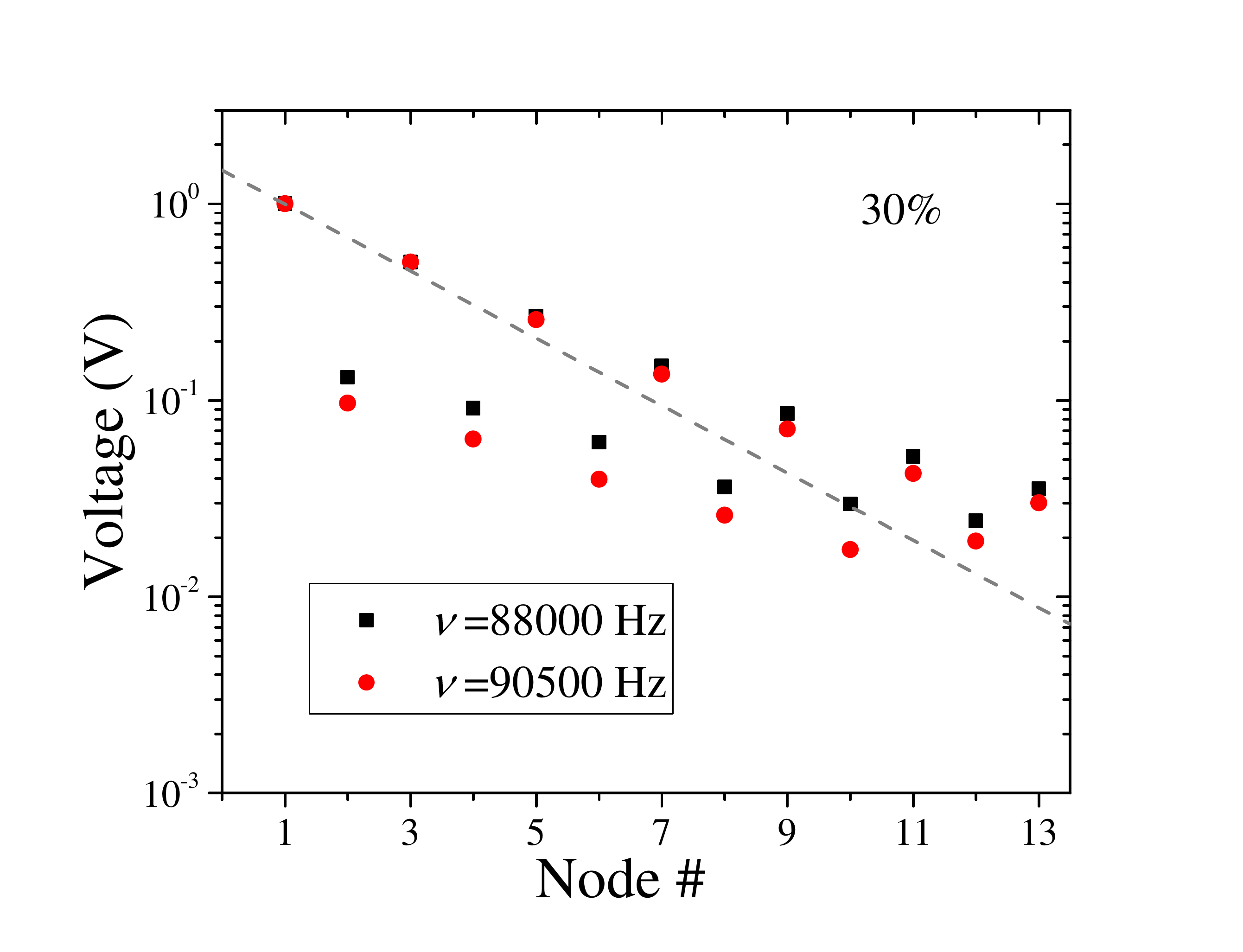}
    \caption{ Memristive SSH circuit simulations with a capacitance distribution of 15\%  and 30\%. (a) Impedance as a function of frequency found averaging 100 random realizations of the circuit. (b) and (c) present the voltage distributions for two selected frequencies.  The parameters of memristive elements are the same as in Fig.~\ref{fig:2}: $R_{on}=10^2$~Ohms and $R_{M}(t=0)=10^4$~Ohms. \label{fig:11}}
\end{figure}

However, note that the chiral-symmetry breaking term is diagonal and depends on $1/R_M$. In fact, we can define its strength by taking the diagonals of the matrix Eq.~(\ref{Btilde}):  $\mathbf{\Delta}=\textrm{diag}(0,\alpha_1,\alpha_2,\cdots)$. Using the projection operators, $P_1$ and $P_2$, we have previously defined, we have $(P_1-P_2)\mathbf{\Delta}(P_1-P_2)=\mathbf{\Delta}$, so the conductance in the presence of memory elements does not respect the full chiral symmetry, but its violation is suppressed by increasing the magnitude of the memristances $R_M$. 

In field-theory language we would say that the symmetry is broken by a ``mass term'' and its strength is proportional to the 
mass itself. This is precisely the definition of {\it custodial symmetry}~\cite{Schwartz}. We thus expect that the delocalized state (\ref{Vdis}) is still protected against perturbations and located in the mid-gap of the continua. 
This is shown explicitly in Fig.~\ref{fig:11}, where we randomly perturb the values of the capacitors. Even up to perturbations 
of $30\%$, the delocalized state is still clearly visible via the peak in the impedance. (See the Supplementary Material for details.)

{\it Conclusions --} In conclusion, we have shown that the concept of custodial symmetry holds also in the classical case. In particular, we 
have used a 1D electrical circuit with memristive elements to emulate the SSH model with memory effects. We have shown both analytically and numerically that memory induces nonlinearities that break chiral symmetry explicitly and spreads the in-gap state across the circuit. Nevertheless, the resulting state is still protected against perturbations due to the promotion of the original chiral symmetry to a 
custodial status. Our predictions, which can be verified experimentally with realistic circuit elements, open up yet another venue to explore some symmetry and topological concepts, which are typically found in quantum systems, in classical settings.

\textit{Acknowledgement --} C. C. C. was supported by the NSF under Grant No. PHY-2011360. M.D. acknowledges financial support from the DOE under Grant No. DE-SC0020892.

\bibliographystyle{apsrev}

\begin{thebibliography}{31}
\bibitem[{IVn()}]{IVnote}
	\bibinfo{note}{The expression differs from that of
		Ref.~\cite{LeeTopolectric18}, where the voltage of node $0$ is included and
		the impedance between node 1 and node $L$ is measured.}
\bibitem[{Non()}]{NonlinearNote}
	\bibinfo{note}{Indeed, there have been studies on nonlinear effects in
		topological systems~\cite{Hadad16,Tukoup20,Saxena20}. However, the lack of
		superposition in nonlinear systems makes it challenging to identify
		topological properties.}
\bibitem[{SSH()}]{SSHEdgeNote}
	\bibinfo{note}{For the SSH model, the edge state survives the addition of terms
		proportional to $\sigma_z$ or $1_2$~\cite{ShenTI} because $V_{loc}$ remains
		an eigenstate but with a shifted eigenvalue.}
\bibitem[{not()}]{notepolarity}
	\bibinfo{note}{Note that the polarity of the memristive elements is not the
		source of the chiral symmetry breaking because a circuit with random
		memristive polarity does not exhibit any qualitative difference.}
\end{thebibliography}




\clearpage
\newpage
\appendix
\section{Supplementary Material: Custodial chiral symmetry in a Su-Schrieffer-Heeger electrical circuit with memory}

\setcounter{page}{1}
\setcounter{figure}{0}
\renewcommand{\thefigure}{S\arabic{figure}}

\subsection{Numerical simulations}

\begin{figure*}[tb]
    \centering
    (a)\includegraphics[width=0.9\columnwidth]{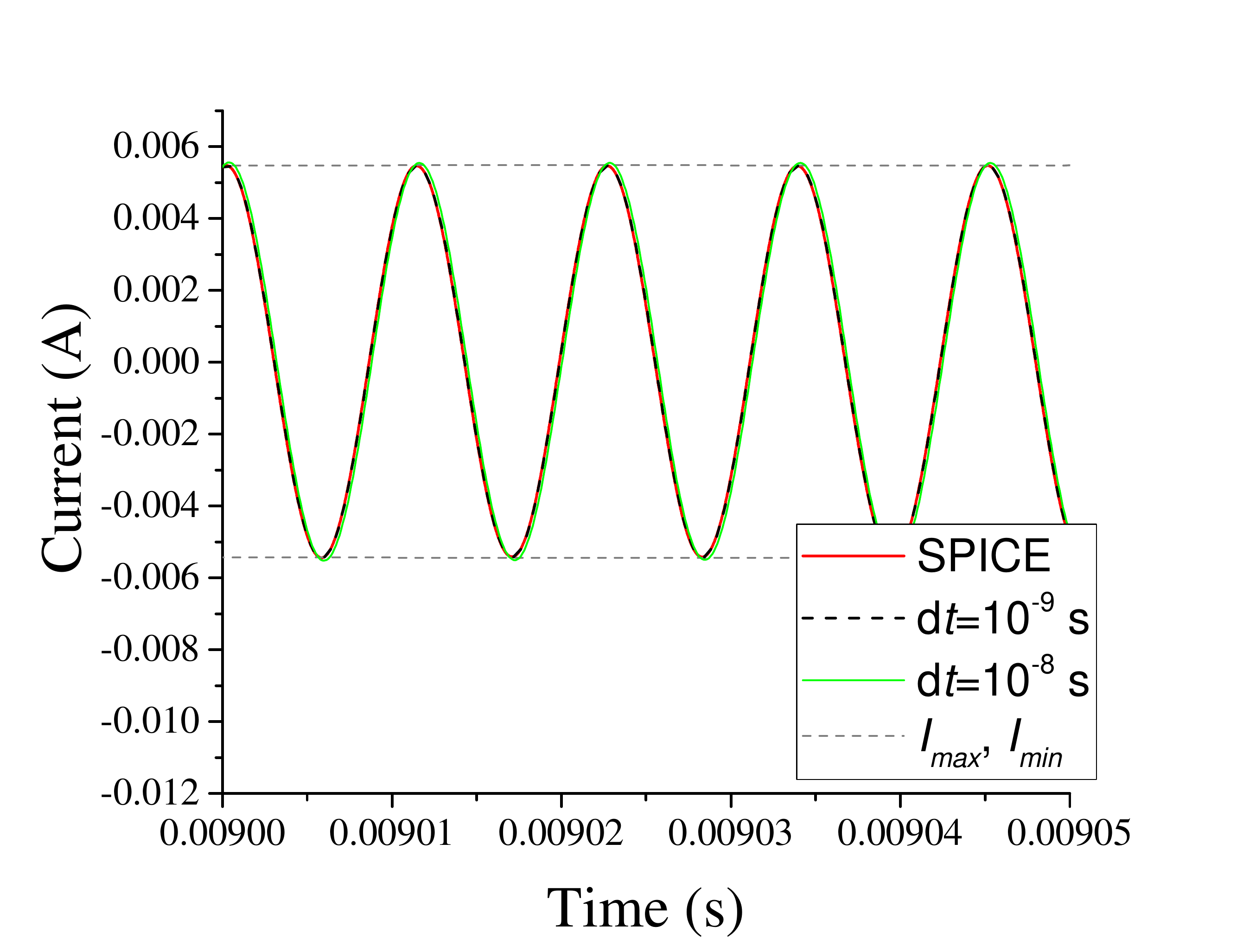} \;
    (b)\includegraphics[width=0.9\columnwidth]{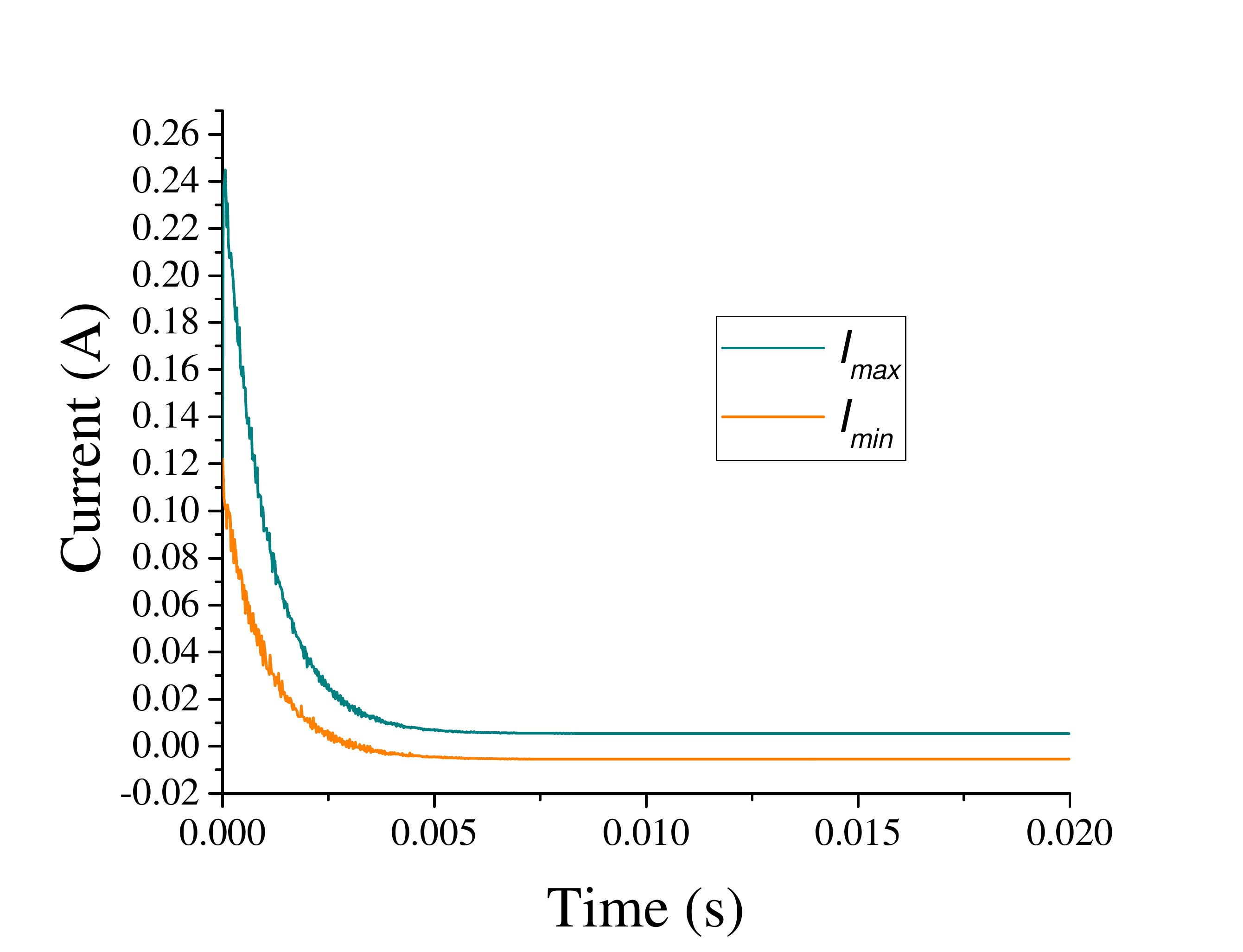} 
    \caption{  (a) Comparison of our custom code and LTspice simulations of Fig.~\ref{fig:1}(a) circuit with $N=4$. (b) Time evolution of  oscillation envelops $I_{max}$ and $I_{min}$. \label{fig:S1}}
\end{figure*}

The numerical integration scheme utilized in this work was based on the Runge–Kutta fourth-order method. We performed a series of tests to check the convergence and correctness of the numerical solution. In particular, we compared results of our simulations  with simulations in LTspice circuit simulator and found an excellent agreement, see Fig.~\ref{fig:S1}(a). This simulation was performed for Fig.~\ref{fig:1}(a)
circuit at $N=4$ using a physical model of inductors (taking into account a small in-series resistance of 10~mOhms). Fig.~\ref{fig:S1}(a) indicates that the integration time step of $10^{-9}$~s is suitable for present simulations.

From time-dependence of the external current $I(t)$, we extracted the current envelops $I_{max}$ and $I_{min}$, see Fig.~\ref{fig:S1}(b). According to this plot, the system reaches the dynamic steady-state at about 10~ms. A similar equilibration time scale was found for other circuits. The current oscillation amplitude was evaluated as  $I_1=(I_{max}-I_{min})/2$.
The ratio of the applied voltage amplitude to the current amplitude $I_1$ was used to estimate the impedance. 

\begin{figure*}[tb]
    \centering
    (a)\includegraphics[width=0.9\columnwidth]{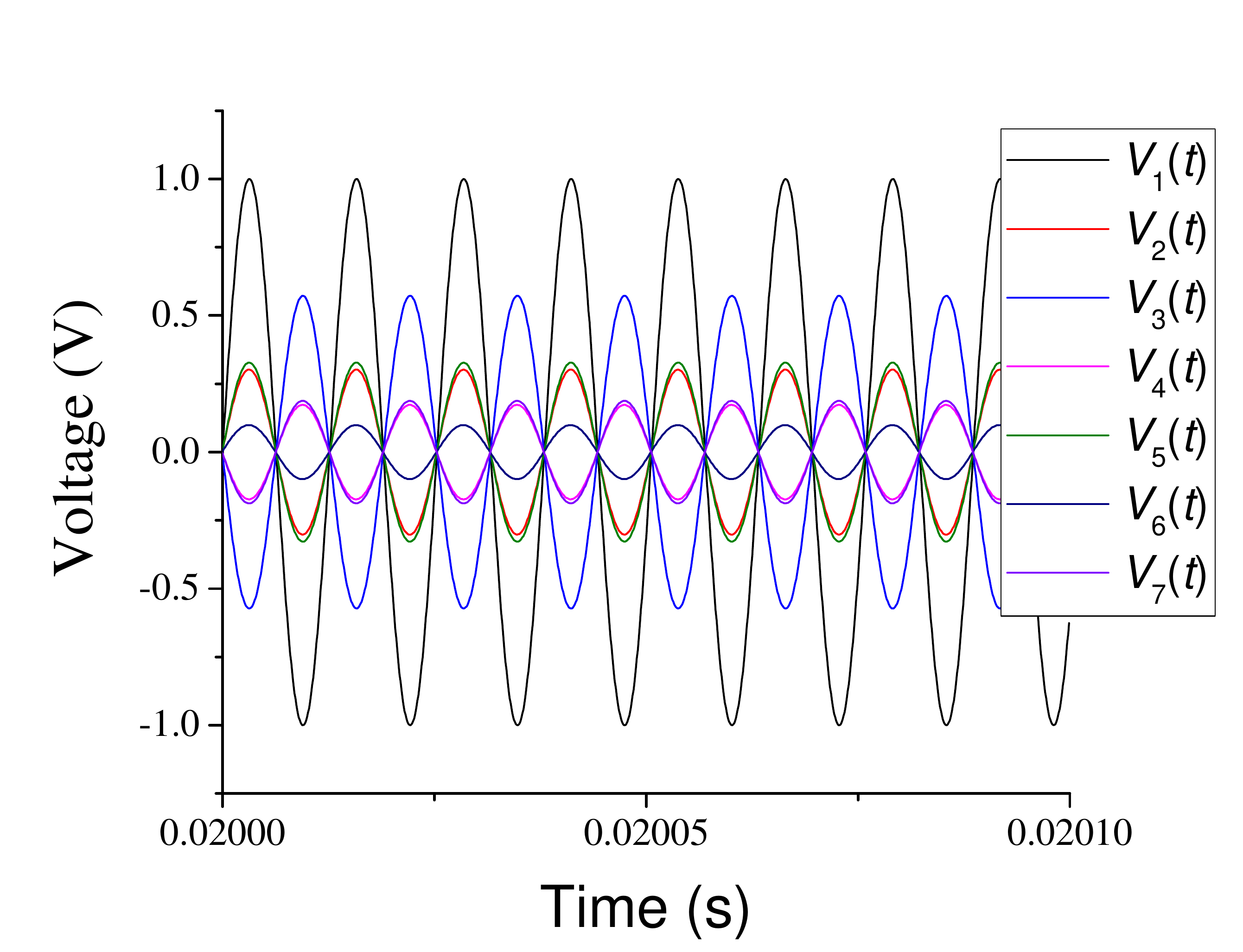} \;
    (b)\includegraphics[width=0.9\columnwidth]{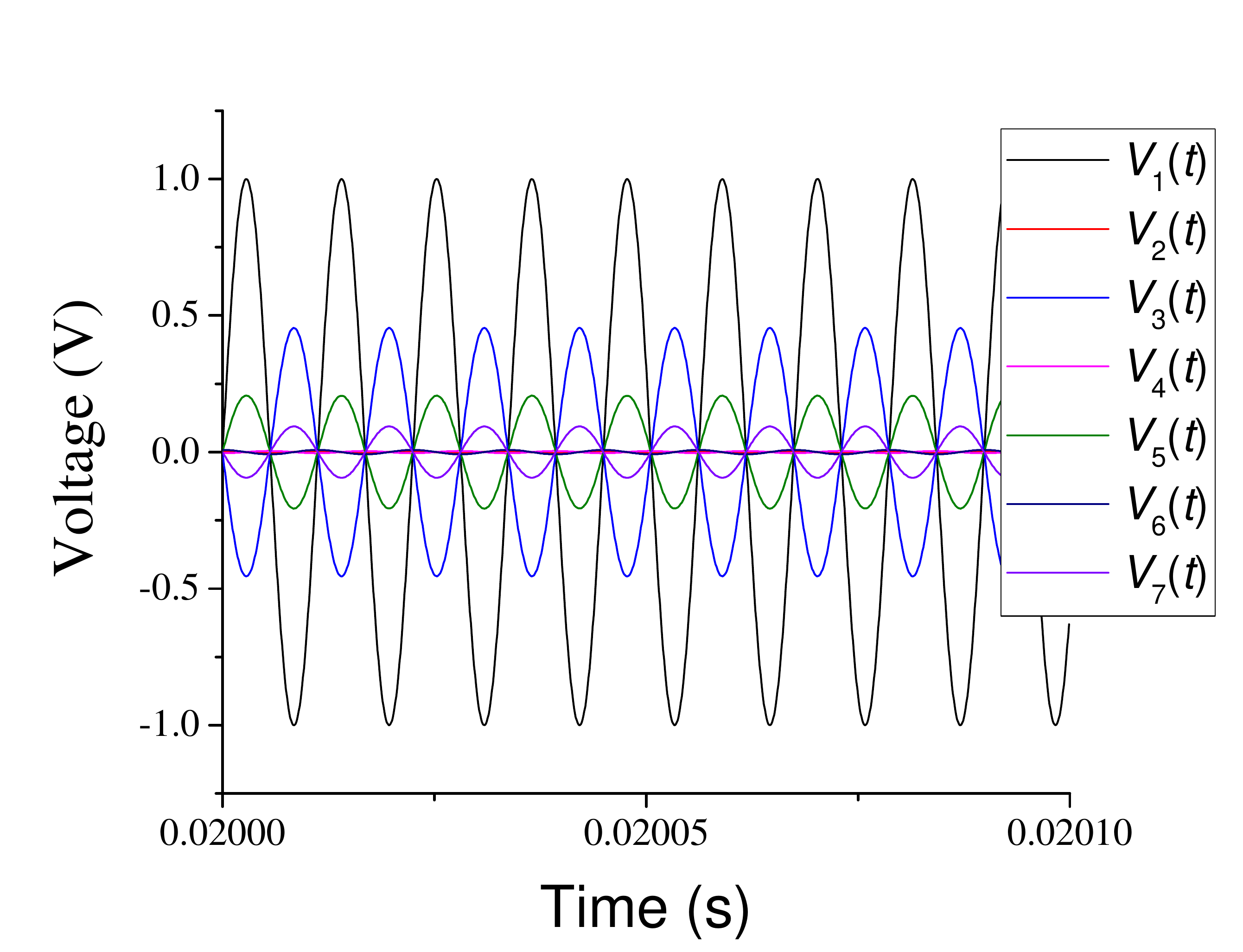}
    \caption{  Voltage waveforms in Fig.~\ref{fig:1}(b) circuit with $N=21$, $R_{on}=10^2$~Ohms, $V_t=0.3$~V at (a) $\nu=79$~kHz, and (b) $\nu=89$~kHz. \label{fig:S3}}
\end{figure*}

\begin{figure*}[tb]
    \centering
    (a)\includegraphics[width=0.9\columnwidth]{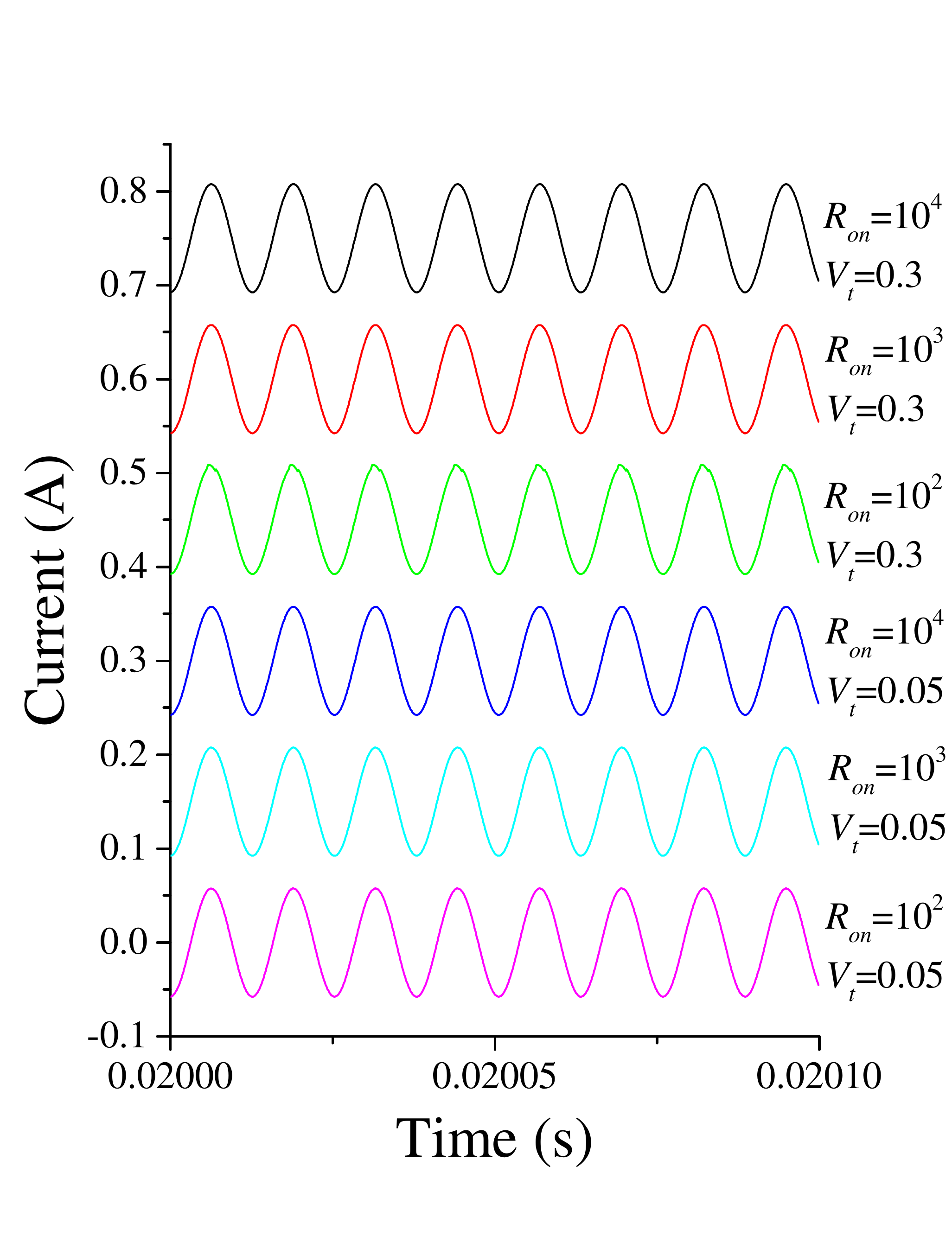} \;
    (b)\includegraphics[width=0.9\columnwidth]{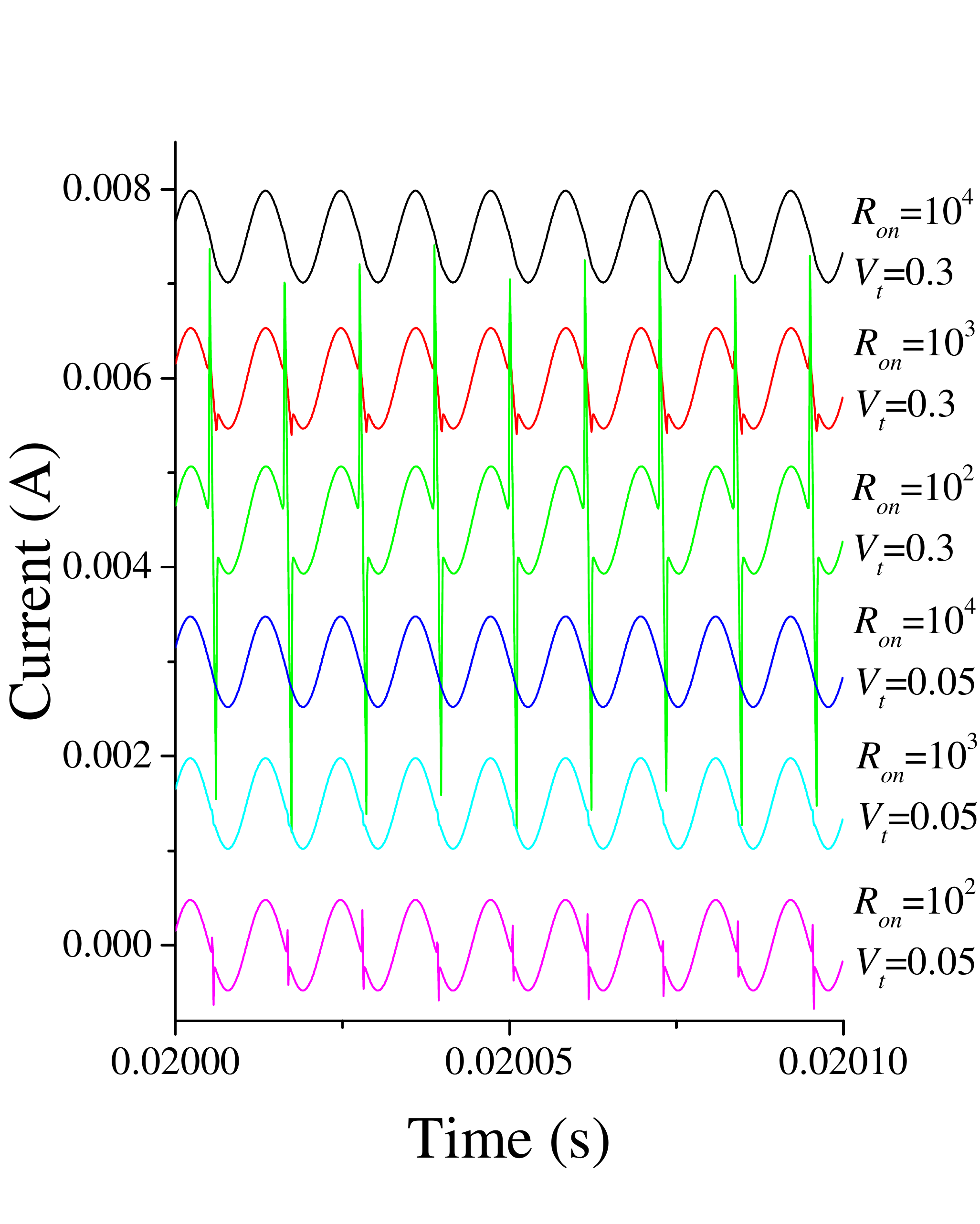}
    \caption{  Current waveforms in Fig.~\ref{fig:1}(b) circuit with $N=21$ at (a) $\nu=79$~kHz, and (b) $\nu=89$~kHz. \label{fig:S2}}
\end{figure*}

We note that in circuits composed of only linear circuit components (such as in Fig.~\ref{fig:1}(a)), perfect sinusoidal current oscillations were observed at long times (see Fig.~\ref{fig:S1}(a)). Examples of voltage waveforms are in Fig.~\ref{fig:S3}. Fig.~\ref{fig:S2} presents examples of current waveform in the memristive SSH circuit. It indicates that in most cases the signal can be approximated well by a sinusoidal waveform. From the point of view of impedance calculation, the deviation is significant only in the case of memristors with a small $R_{on}=10^2$~Ohms and larger threshold voltage $V_t=0.3$~V. Physically, the spikes in Fig.~\ref{fig:S2}(b) are due to the switching of the memristor connecting nodes 0 and 1 into $R_{on}$.

\begin{figure*}[tb]
    \centering
    (a)\includegraphics[width=0.9\columnwidth]{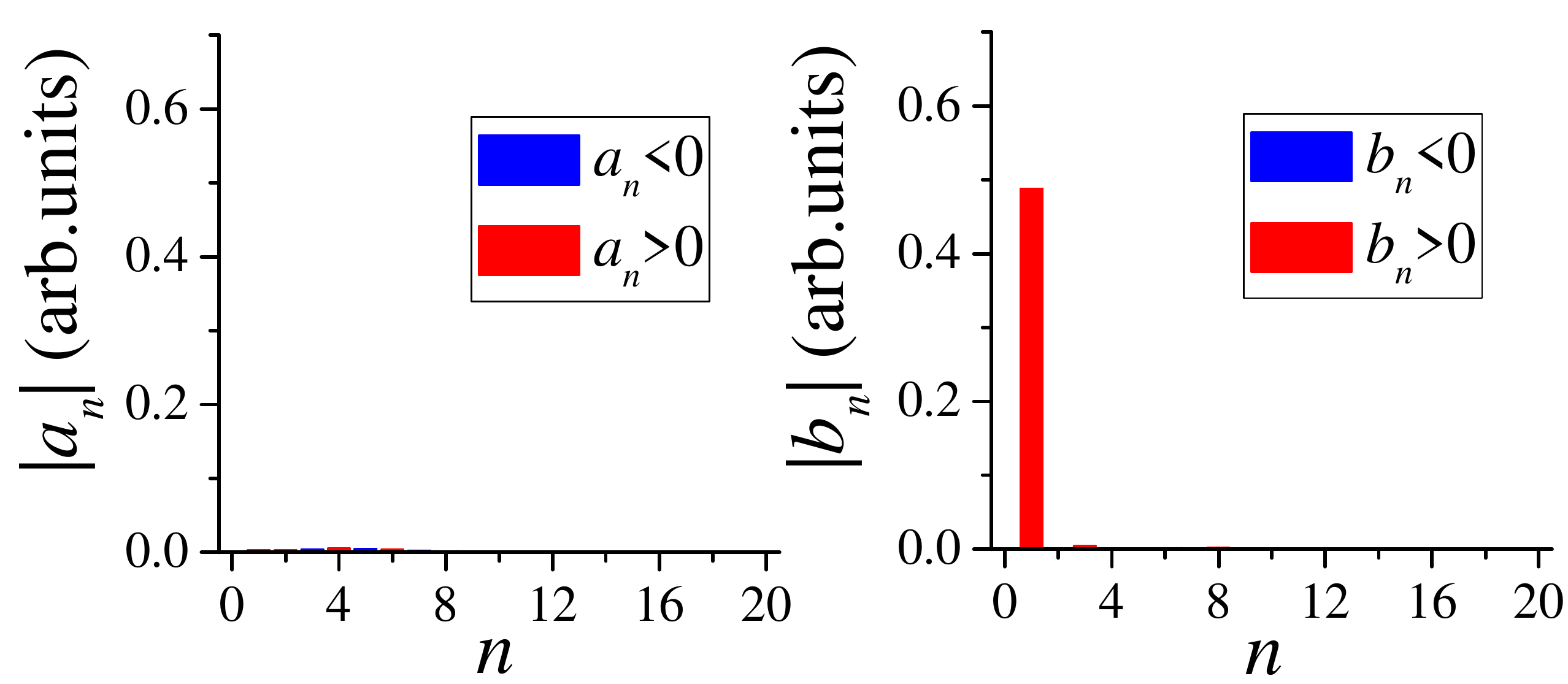} \;
    (b)\includegraphics[width=0.9\columnwidth]{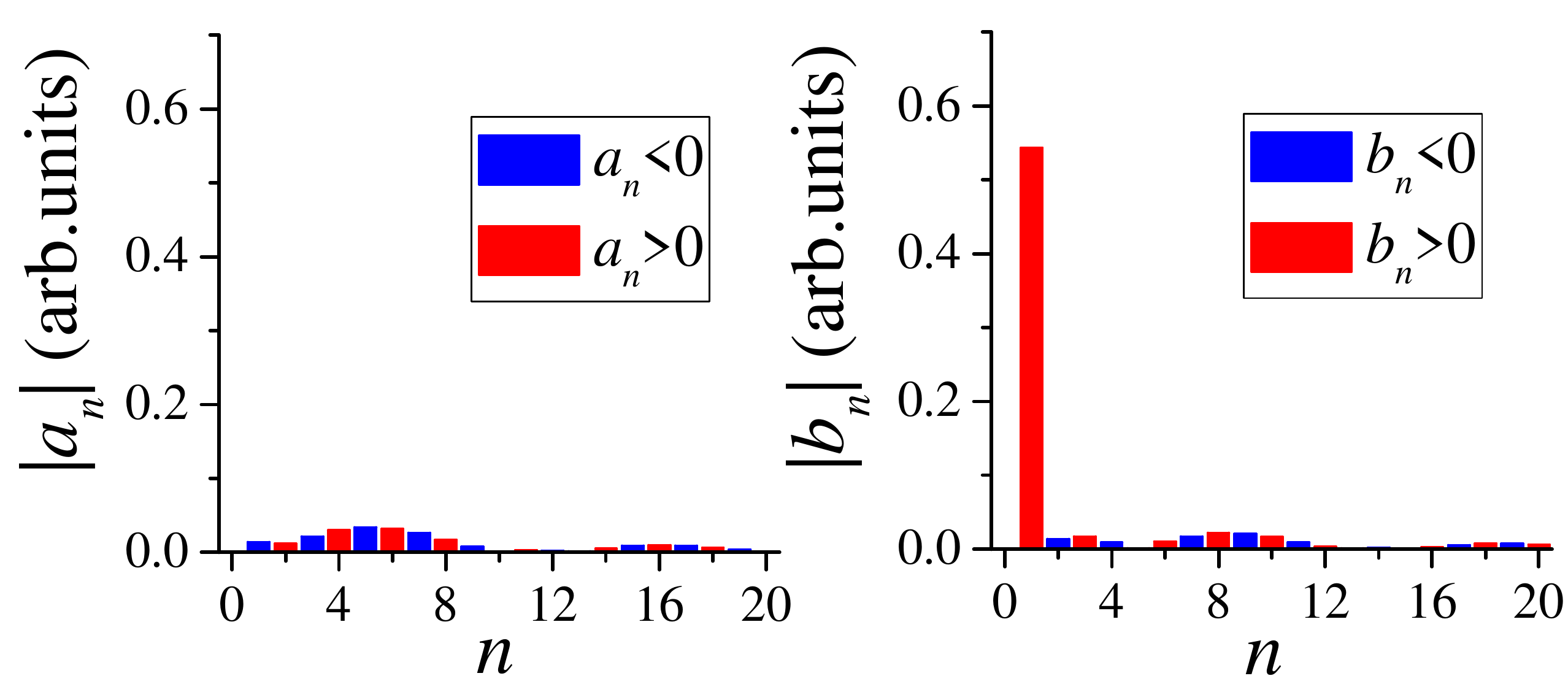} 
    (c)\includegraphics[width=0.9\columnwidth]{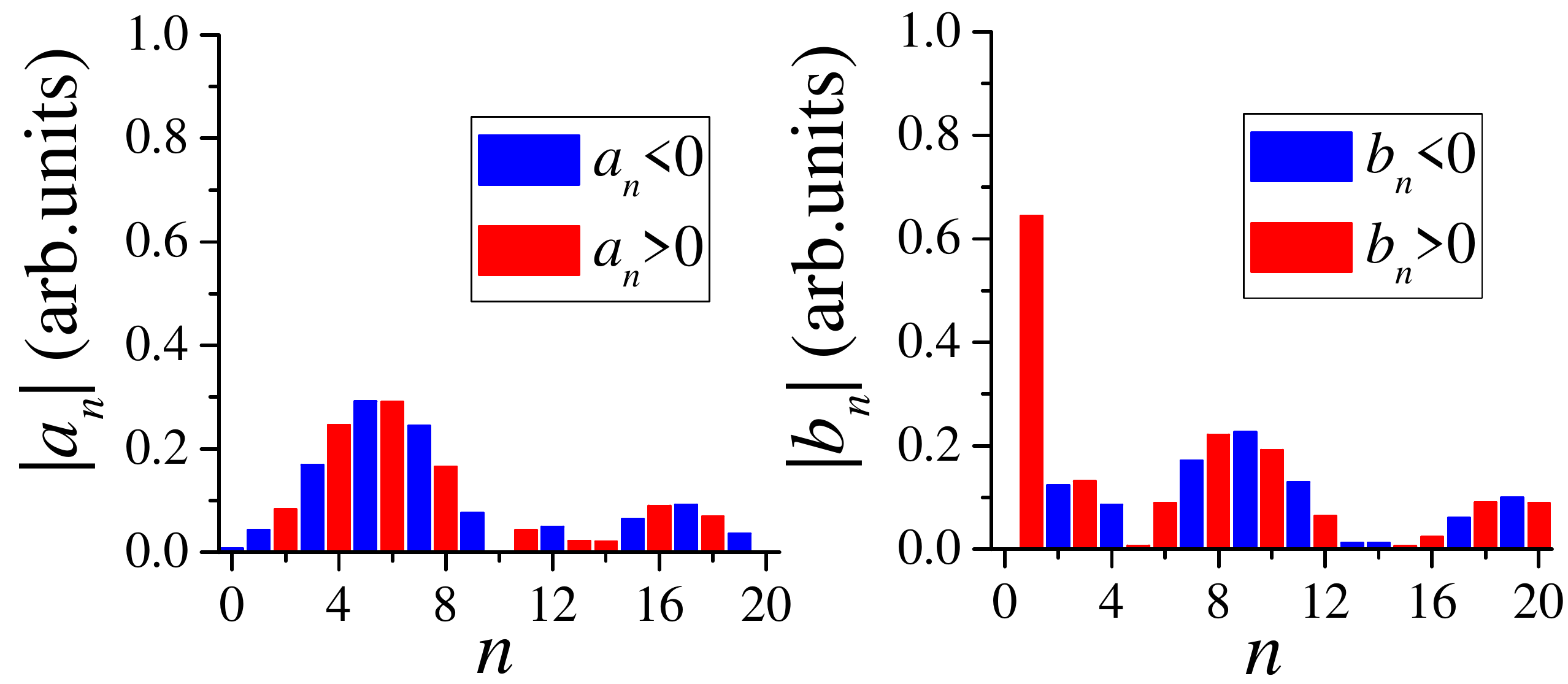} \;
    (d)\includegraphics[width=0.9\columnwidth]{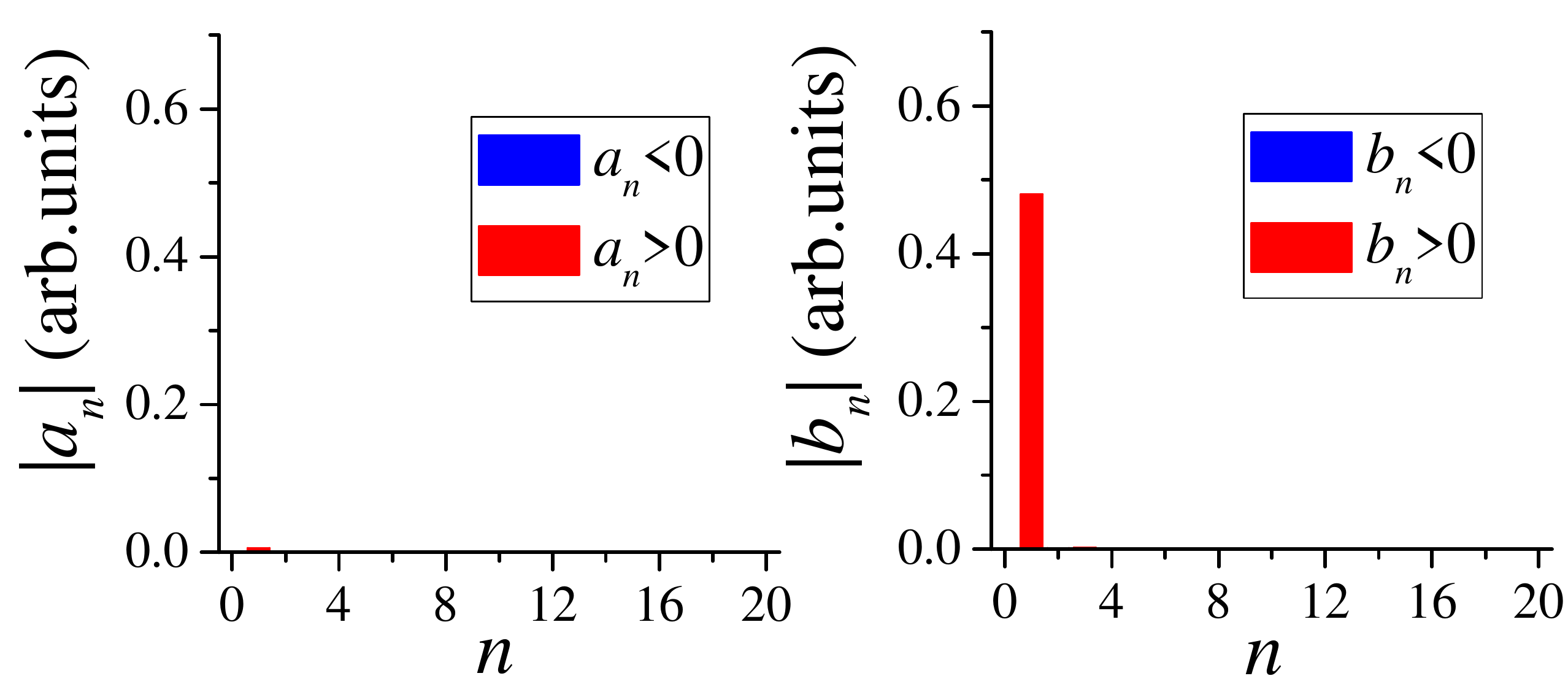}
    (e)\includegraphics[width=0.9\columnwidth]{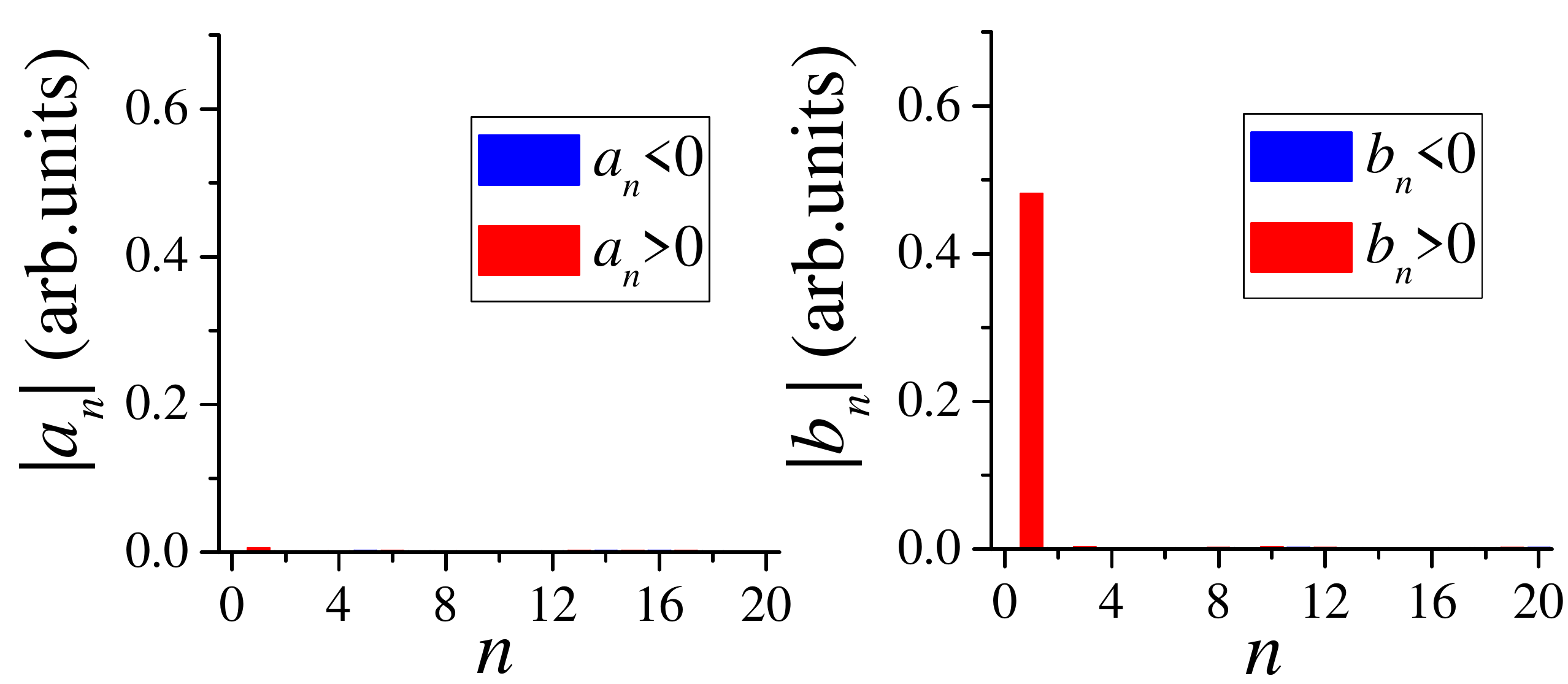} \;
    (f)\includegraphics[width=0.9\columnwidth]{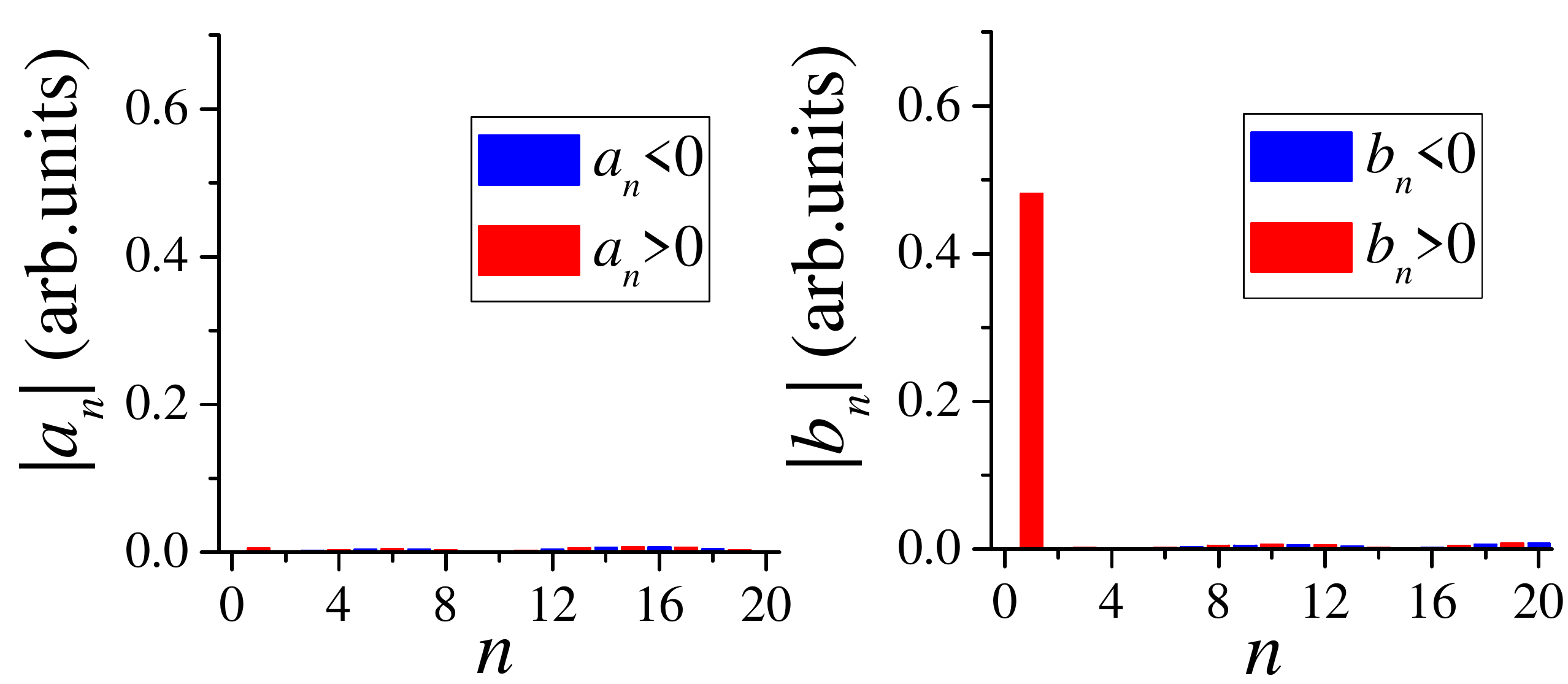}
    \caption{Fourier series coefficients for the current waveforms in Fig. \ref{fig:S2}(b). (a) to (f) correspond to the top to bottom curves in Fig. \ref{fig:S2}(b). Here, $a_n$-s and $b_n$-s are the standard cosine and sine coefficients, respectively. The sign of coefficient is indicated by the color. \label{fig:S2229}}
\end{figure*}

The quantities presented in the main text demonstrate the steady-state oscillations in the circuit. To extract those quantities, we skipped the initial transient evolution interval and identified the amplitudes of oscillations for node voltages and external current. To support the application of the approximate Fourier analysis to the memristive circuit, Fig.~\ref{fig:S2229} shows the Fourier series coefficients of Fig. \ref{fig:S2}. In all but one case of low $R_{on}$ (Fig.~\ref{fig:S2229}(c)), there is a dominant component with the weight of higher modes decaying away, justifying the approximate Fourier analysis in the steady states.

\subsection{Regular resistors in series with capacitors}
\begin{figure*}[tb]
    \centering
    (a)\includegraphics[width=0.9\columnwidth]{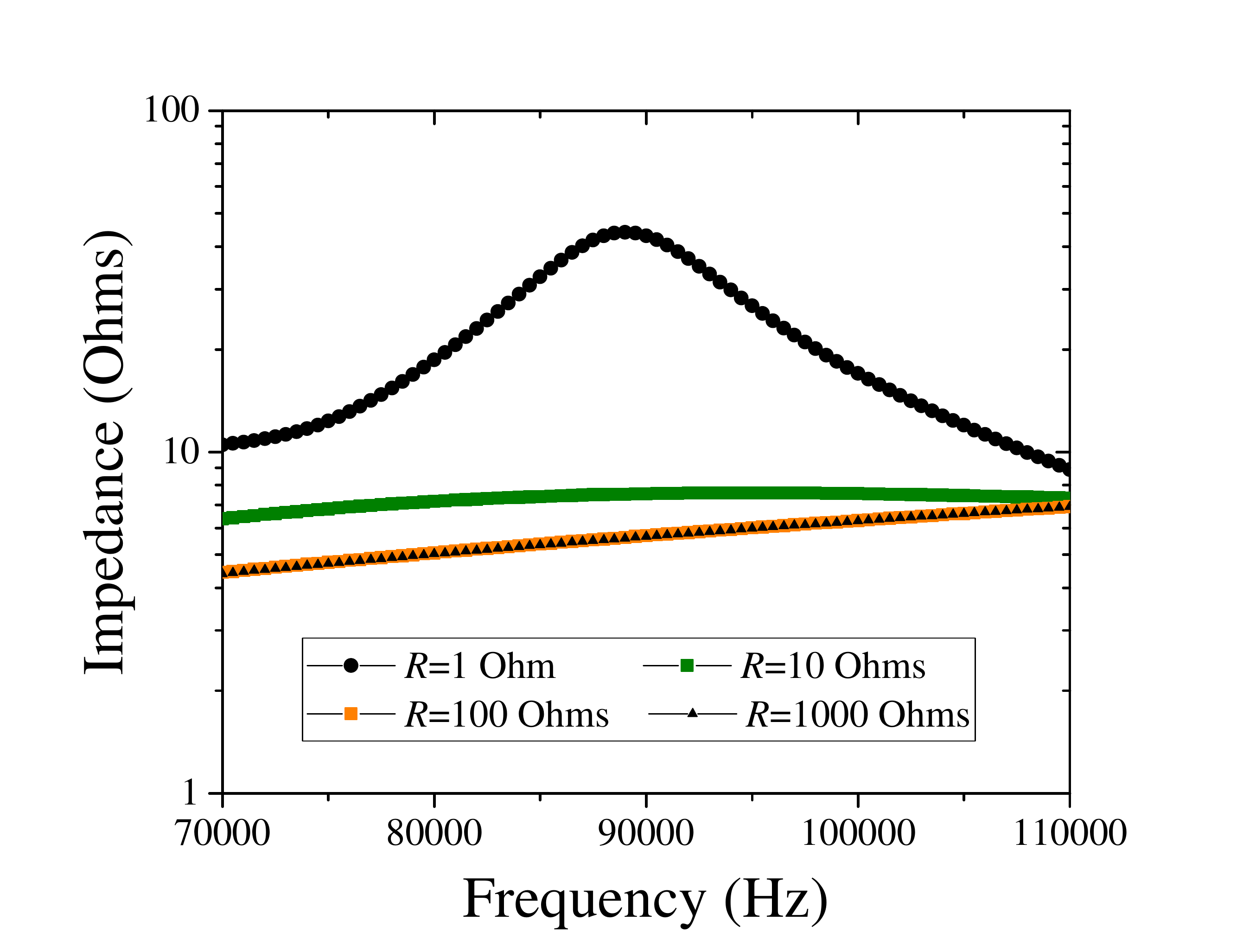} \;
    (b)\includegraphics[width=0.9\columnwidth]{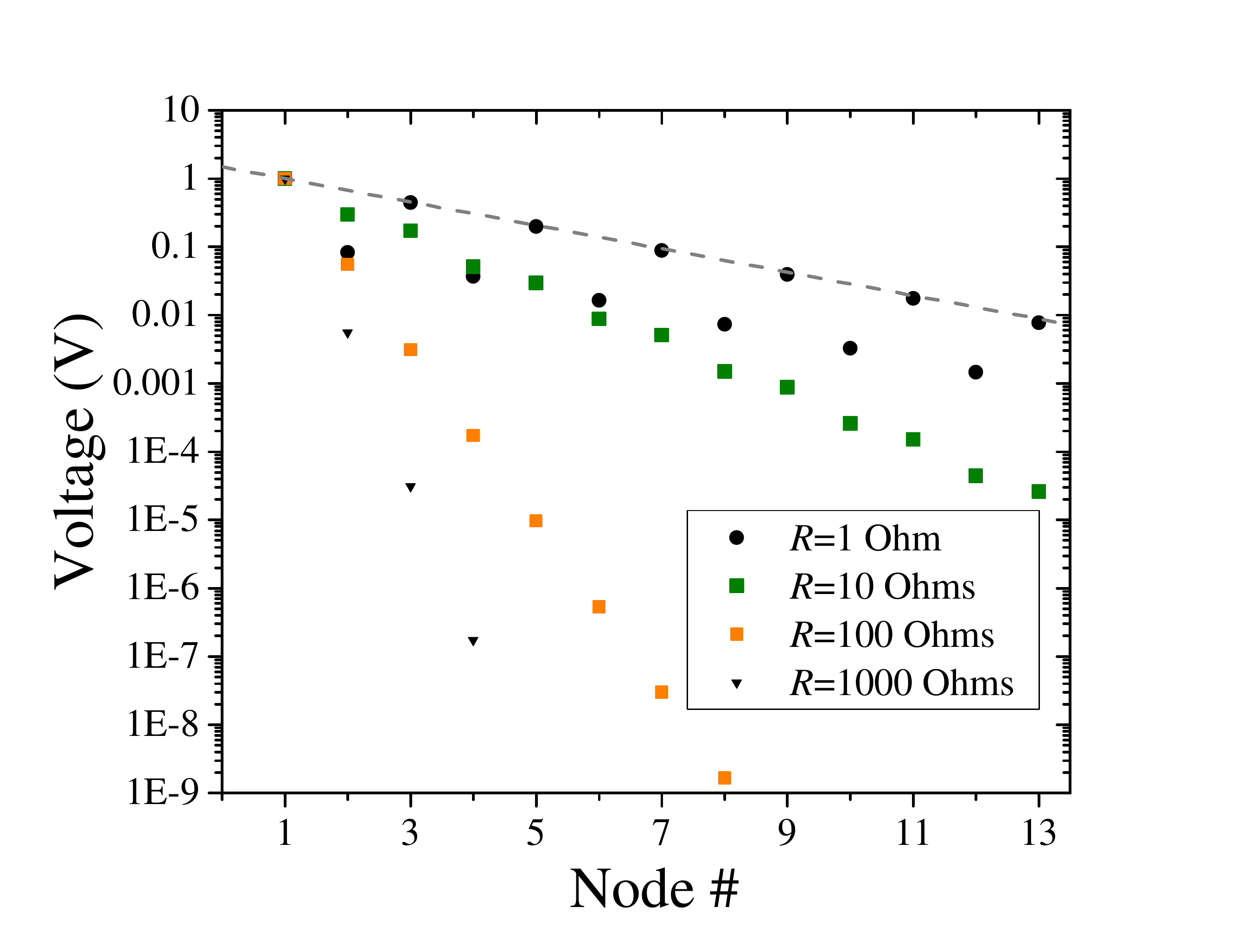}\\
    (c)\includegraphics[width=0.9\columnwidth]{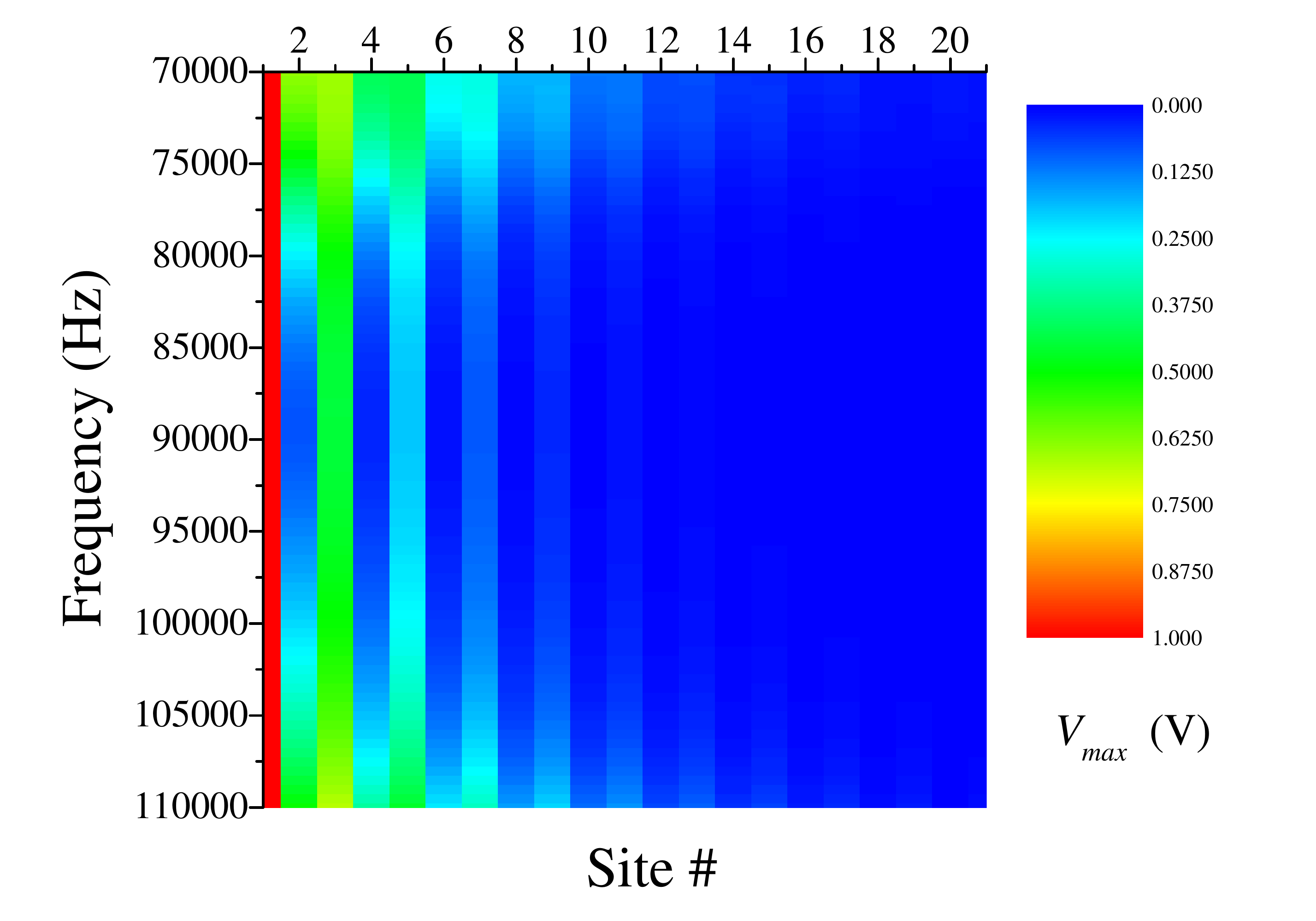} \;
    (d)\includegraphics[width=0.9\columnwidth]{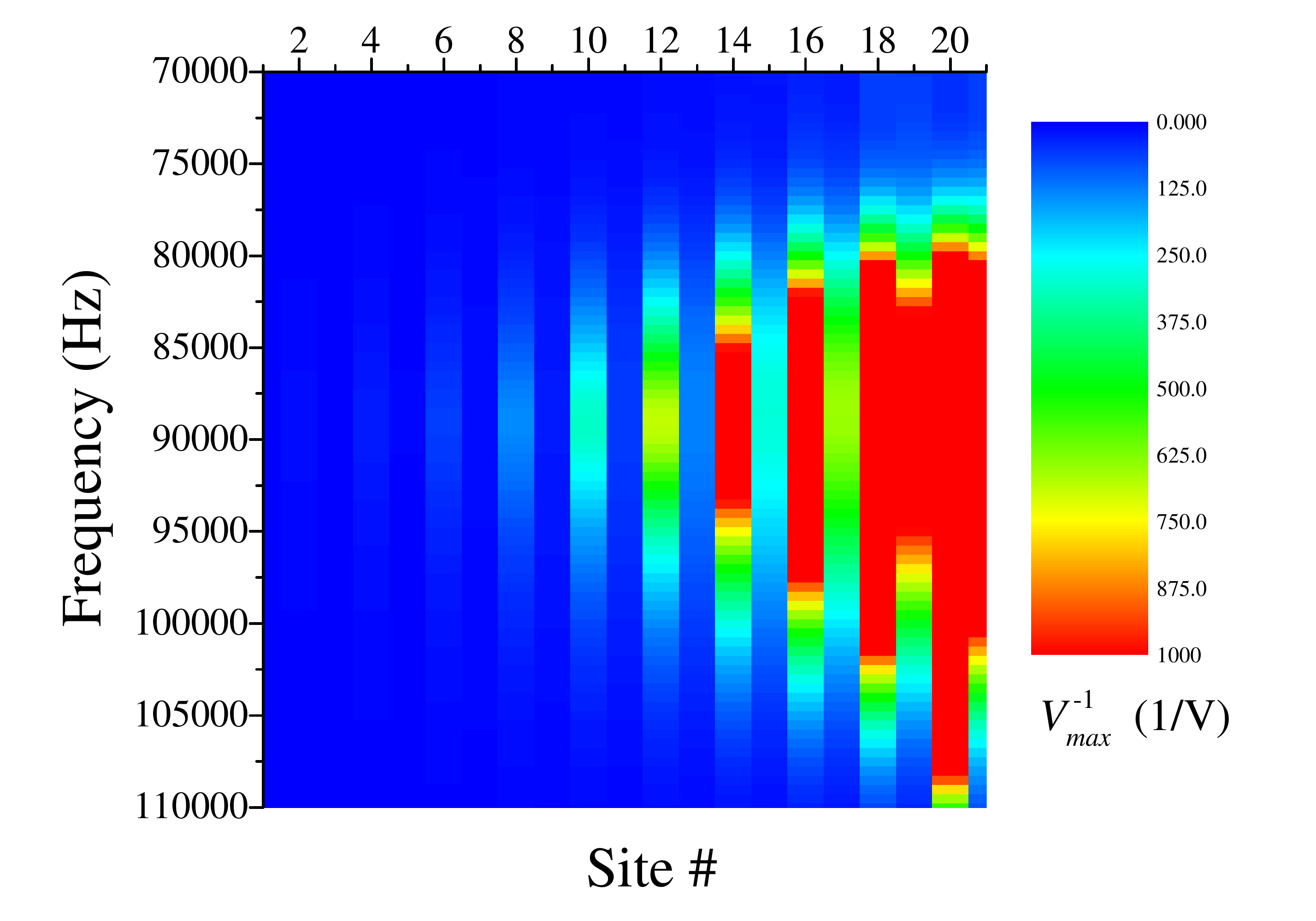}\\
    \caption{ SSH circuit simulations with resistors connected in-series with capacitors. (a) Impedance as a function of frequency. (b) Voltage distributions at 89~kHz. (c), (d) Voltage and inverse voltage distributions plots for $R=1$~Ohm. \label{fig:SM1}}
\end{figure*}

If we place regular resistors in series with the capacitors, the chiral-symmetry protected mid-gap state is still visible provided the 
resistance is not too large so as to close the current flow in the circuit. This is because the resistors in series with the capacitors 
provide the current at each node of the type $I=V/[(i\omega C)^{-1} + R]=[i\omega C/(1+i\omega C R) ]V$, with the current going to zero with increasing $R$. This effect on the mid-gap state is clearly seen in Fig.~\ref{fig:SM1}.\\

\subsection{Voltage profiles and extended plot corresponding to Fig. 2}
\begin{figure*}[tb]
    \centering
    (a)\includegraphics[width=0.9\columnwidth]{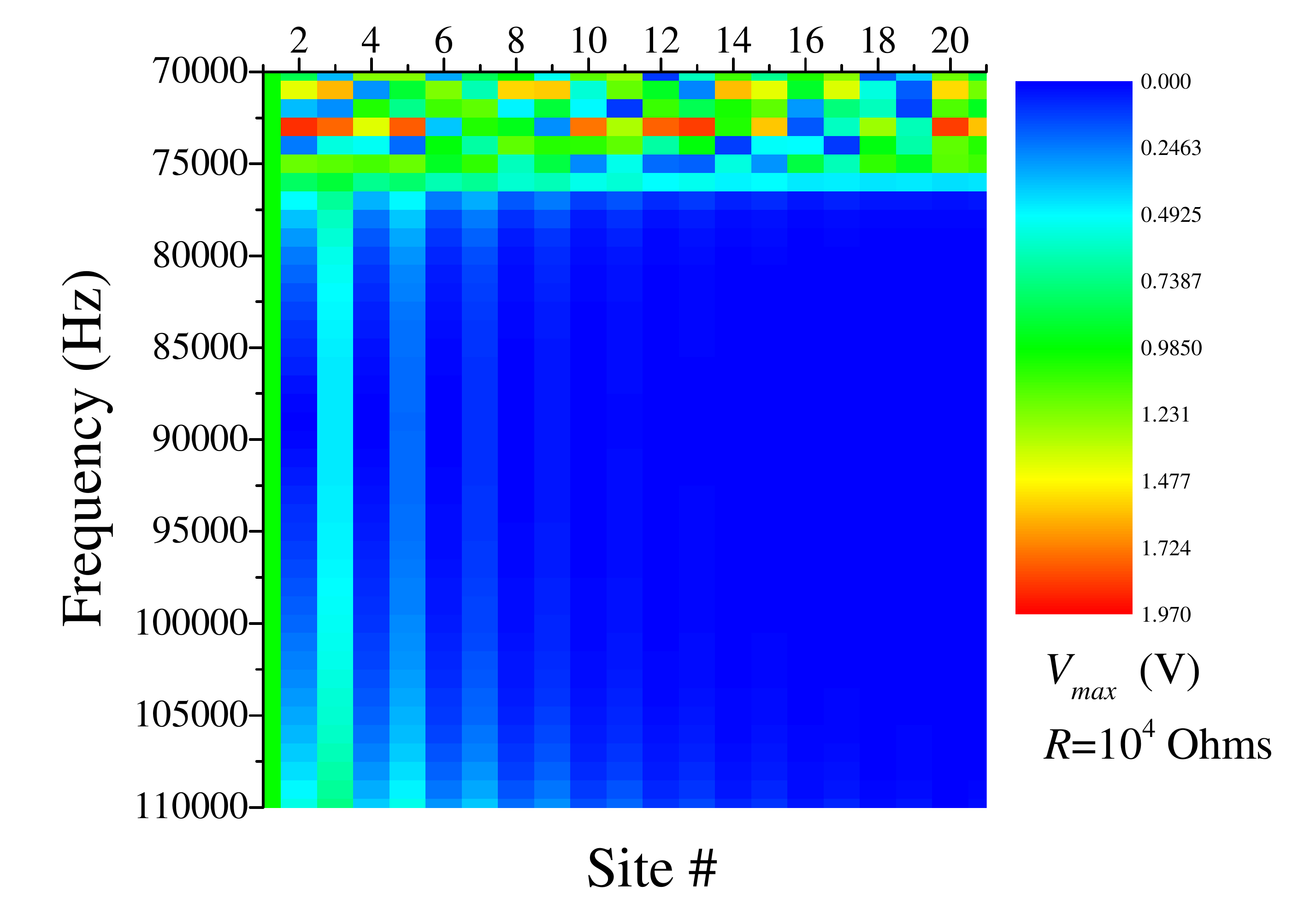} \;
    (b)\includegraphics[width=0.9\columnwidth]{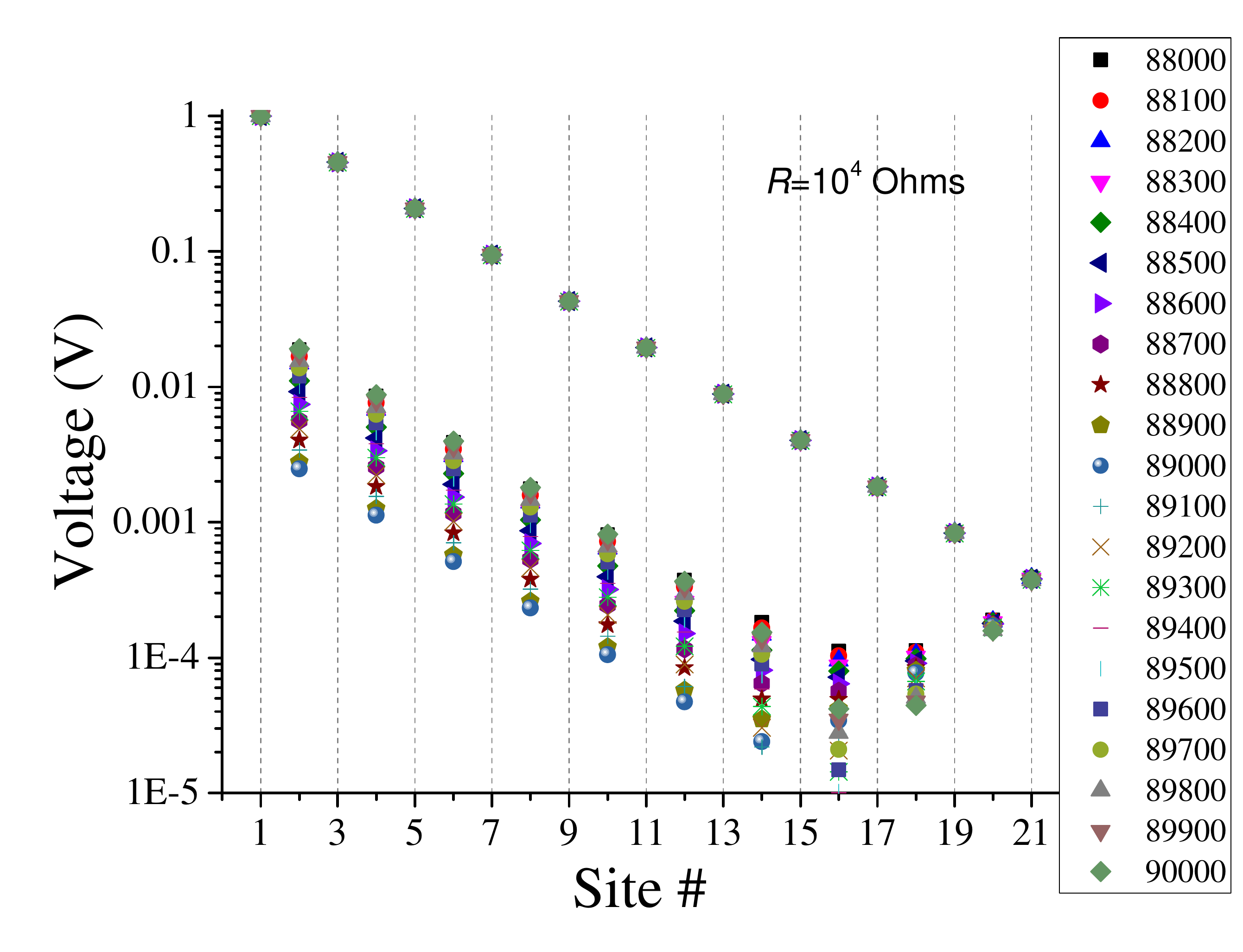} \\
    (c)\includegraphics[width=0.9\columnwidth]{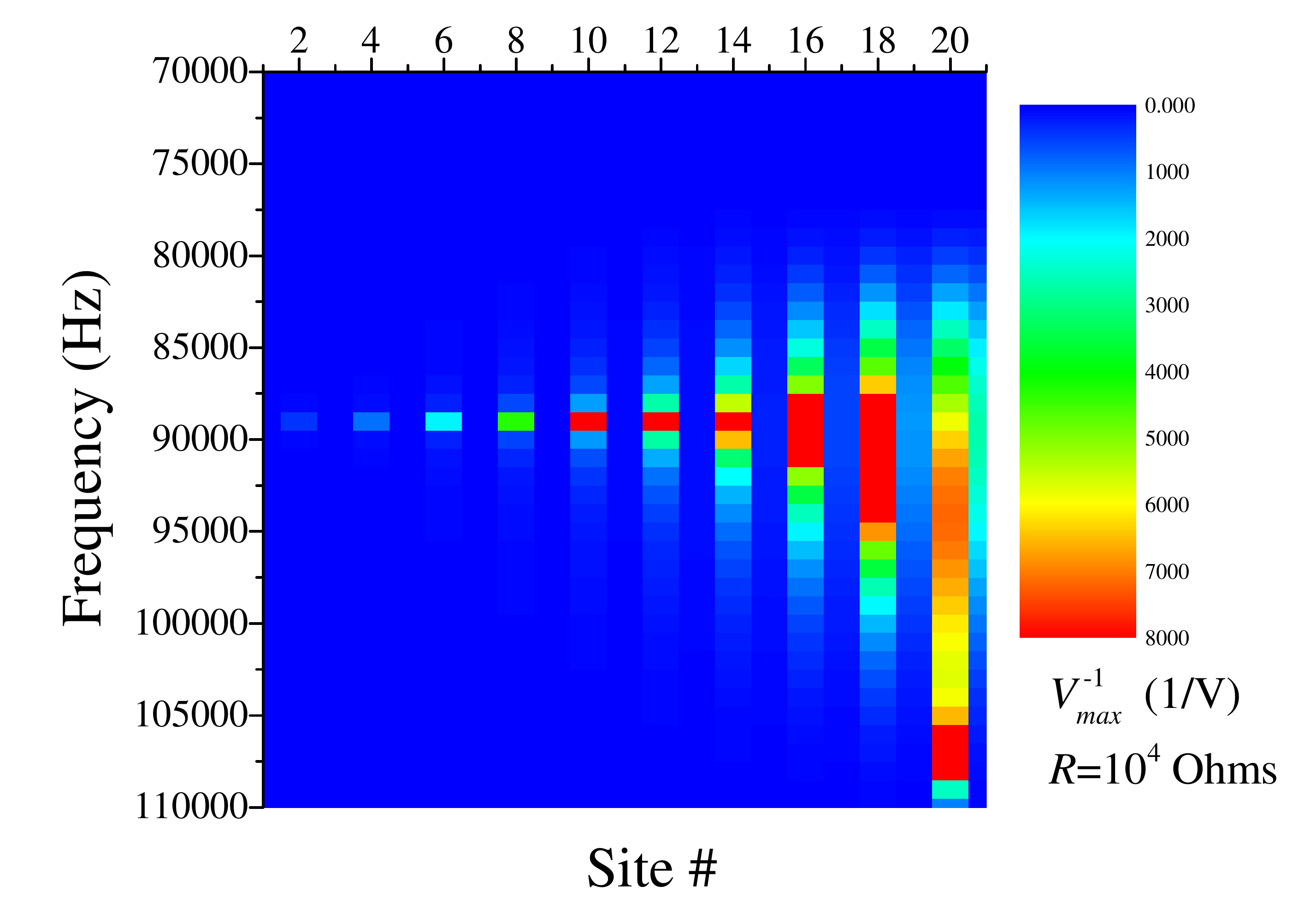}\;
    (d)\includegraphics[width=0.9\columnwidth]{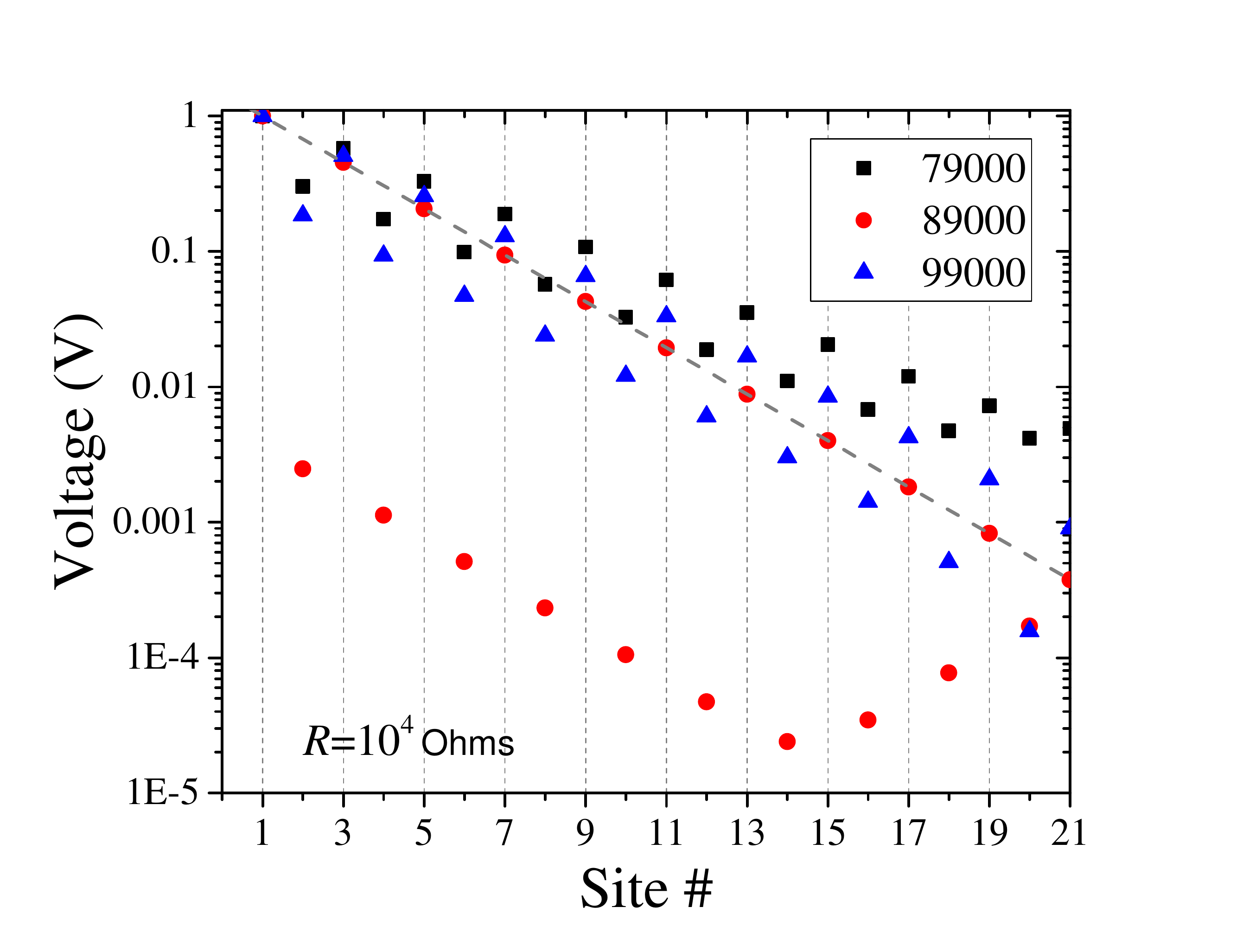}
    \caption{ Another representation of Fig.~\ref{fig:2} with $R=10^4$~Ohms. The edge state can not be clearly distinguished in (a) as its main feature is the suppression of voltages at even sites (for different frequencies), see (b). However, the edge state can be distinguished in the inverse voltage plot, (c). (d) Comparison of voltage distributions in the edge state (red circles) and outside (black squares and blue triangles). This plot indicates the exponential decay in all three cases. The distinguishing feature of the edge state is that the voltages at even sites are very low.  \label{fig:220}}
\end{figure*}

\begin{figure*}[tb]
    \centering
    \includegraphics[width=0.9\columnwidth]{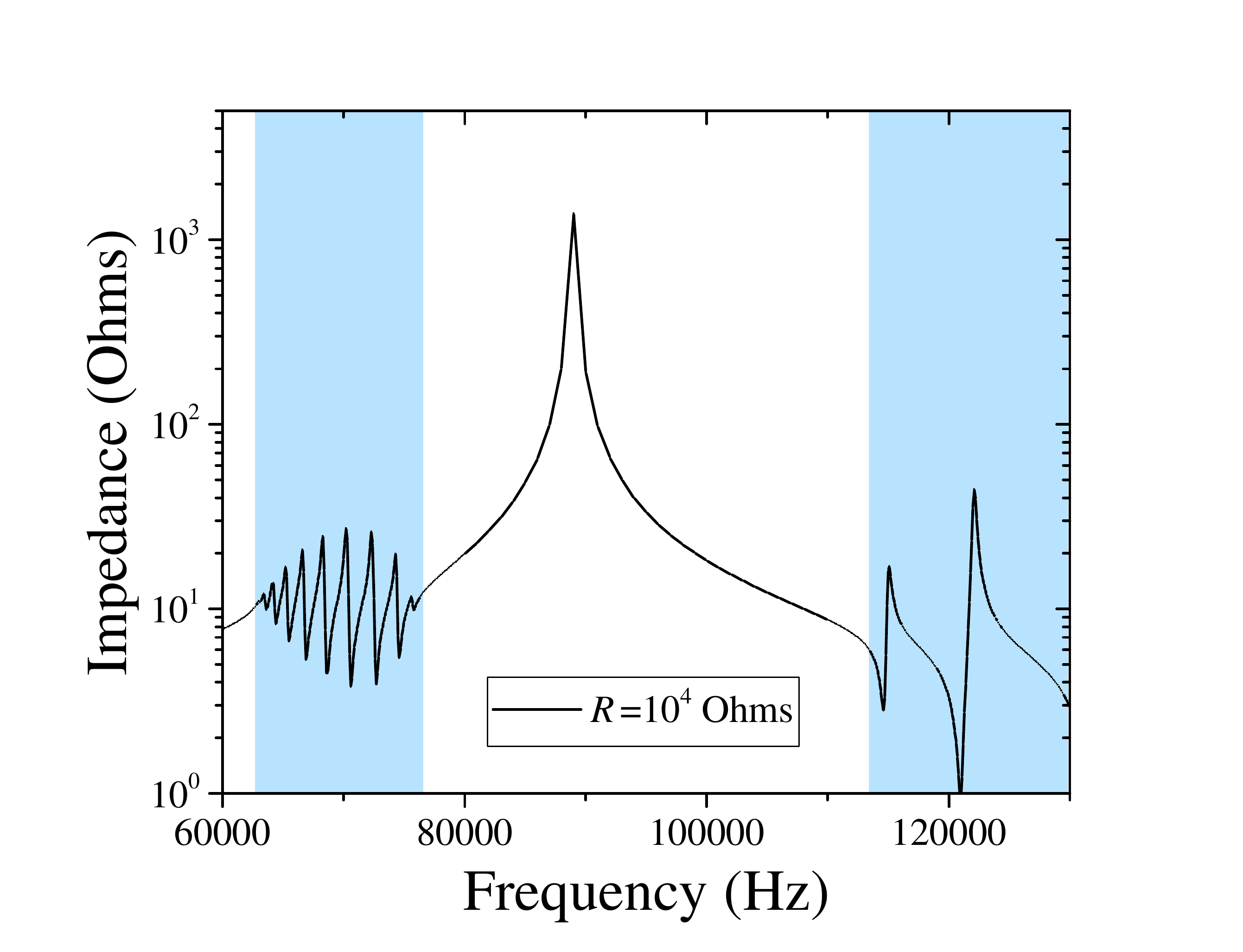} 
    \caption{Extended frequency plot corresponding to the $R=10^4$~Ohms curve in Fig.~\ref{fig:2}(a) of the main text. The contributions of the SSH band states are highlighted in shaded color. \label{fig:S2220}}
\end{figure*}

The voltage profiles corresponding to Fig. 2 in the main text are shown in Fig.~\ref{fig:220}. Moreover, a plot of the impedance in an extended frequency range shows the contributions from the band states of the admittance matrix on the two sides of the localized-state peak, as shown in Fig.~\ref{fig:S2220}.

\subsection{Memristive elements in series with capacitors}
If the memristive elements are in series with the capacitors, the expression for a single node becomes $I=V/[(i\omega C)^{-1} + R_M]=[i\omega C/(1+i\omega C R_M) ]V$. In the resulting SSH memcircuit, the conductance matrix can be obtained by replacing $C_{1,2}$ in Eq.~\eqref{eq:JSSH} of the main text with $\hat{C}_{1,2}=[C_{1,2}/(1+i\omega C_{1,2}R_{M,\{1,2\}})]$. The conductance matrix then becomes complex-valued and the equations are nonlinear. Therefore, the same arguments as the ones we made in the main text for the memristive elements in parallel with the capacitors hold. 

The admittance matrix now becomes $\mathbf{G}=\hat{A}\mathbf{1}_{N}+\hat{\mathbf{B}}$. Here $\hat{A}(x_j)=i\omega[-(\omega^2 L(x_j)^{-1}+\hat{C}_L(x_{j-1})+\hat{C}_R(x_{j+1})]$, where $\hat{C}_L$ and $\hat{C}_R$ are the capacitor to the left and to the right of the node, and 
\begin{equation}\label{Btilde_s}
\hat{\mathbf{B}}=\left(\begin{array}{ccccc}
0 & \hat{C}_2(x_2) & 0 & 0 &\cdots \\
\hat{C}_2(x_1) & \alpha_1 & \hat{C}_1(x_3) & 0 &\cdots \\
0 & \hat{C}_1(x_2) & \alpha_2 & \hat{C}_2(x_4) & \cdots \\
\vdots & \vdots & \vdots & \vdots & \ddots
\end{array}
\right).
\end{equation}
Here, the $\alpha_j=\hat {A}(x_{j+1})-\hat{A}(x_1)$. 

If $R_{M,\{1,2\}}\gg 1$, we may replace $\hat{C}_{1,2}$ by $i \omega/R_{M,\{1,2\}}$. Then $\hat{B}$ has the form of the SSH model with alternating off-diagonal elements. The nonlinearity terms on the diagonal, however, still break the symmetry.

\begin{figure*}[tb]
    \centering
    (a)\includegraphics[width=0.9\columnwidth]{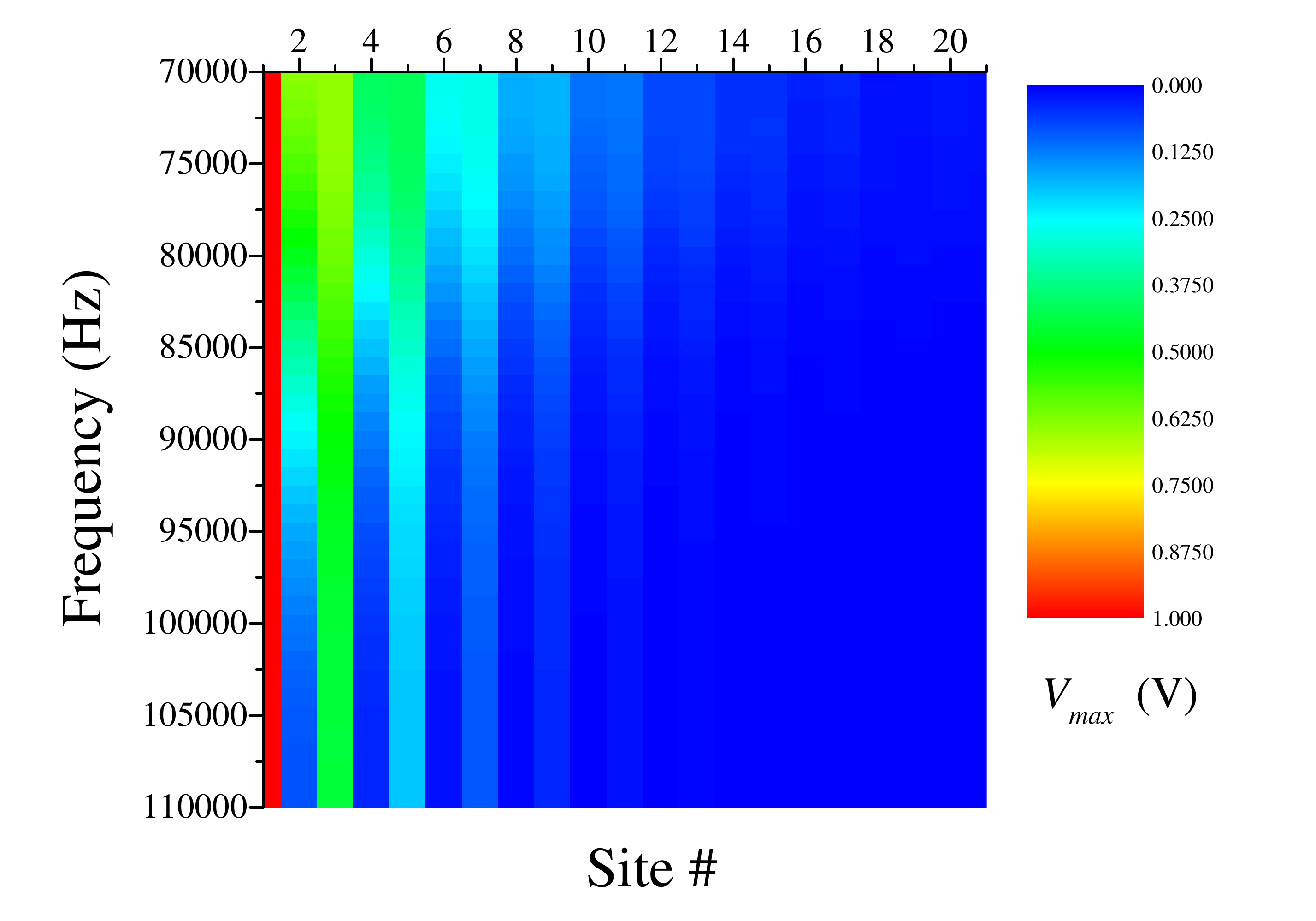} \;
    (b)\includegraphics[width=0.9\columnwidth]{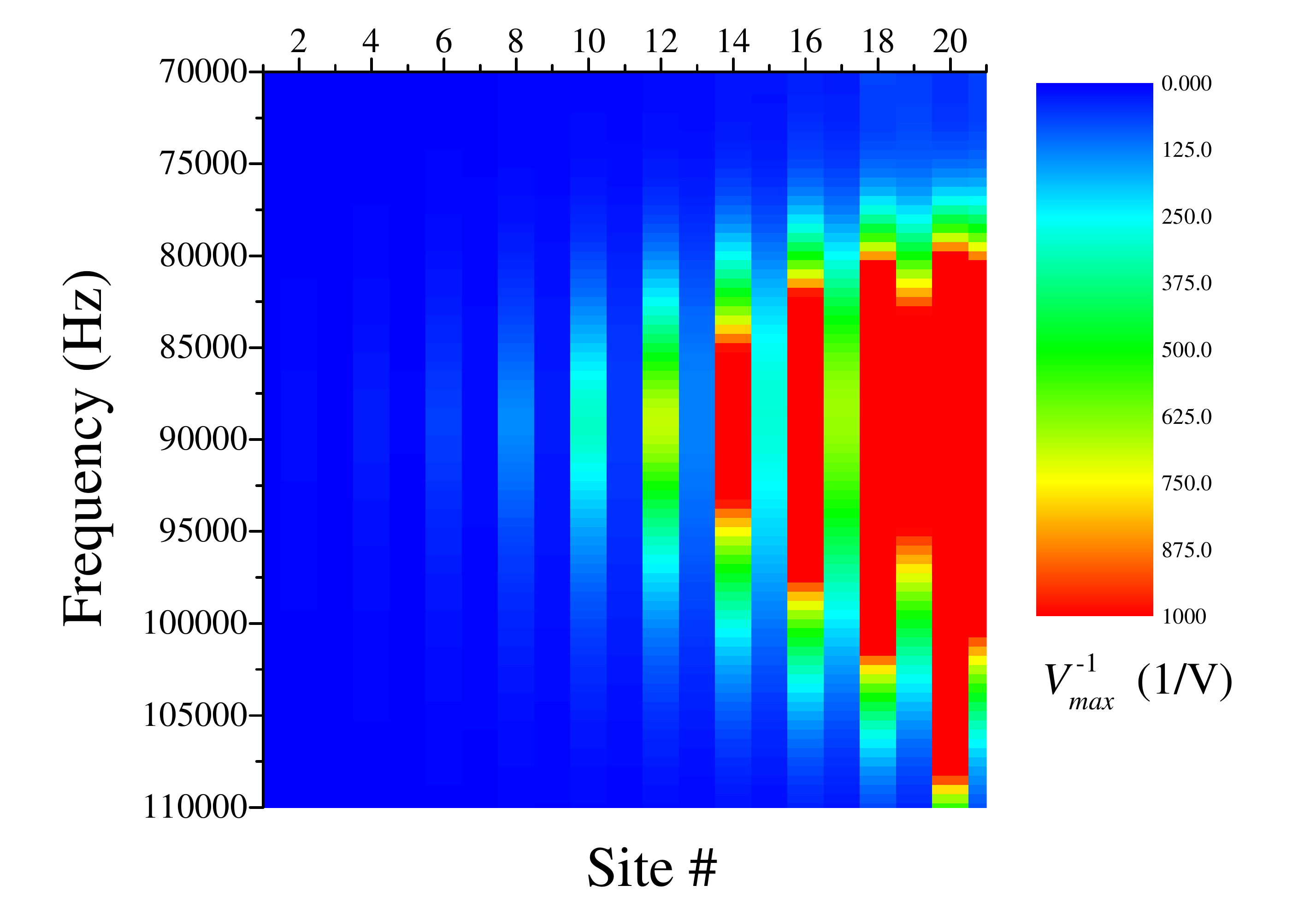} \\
    (c)\includegraphics[width=0.9\columnwidth]{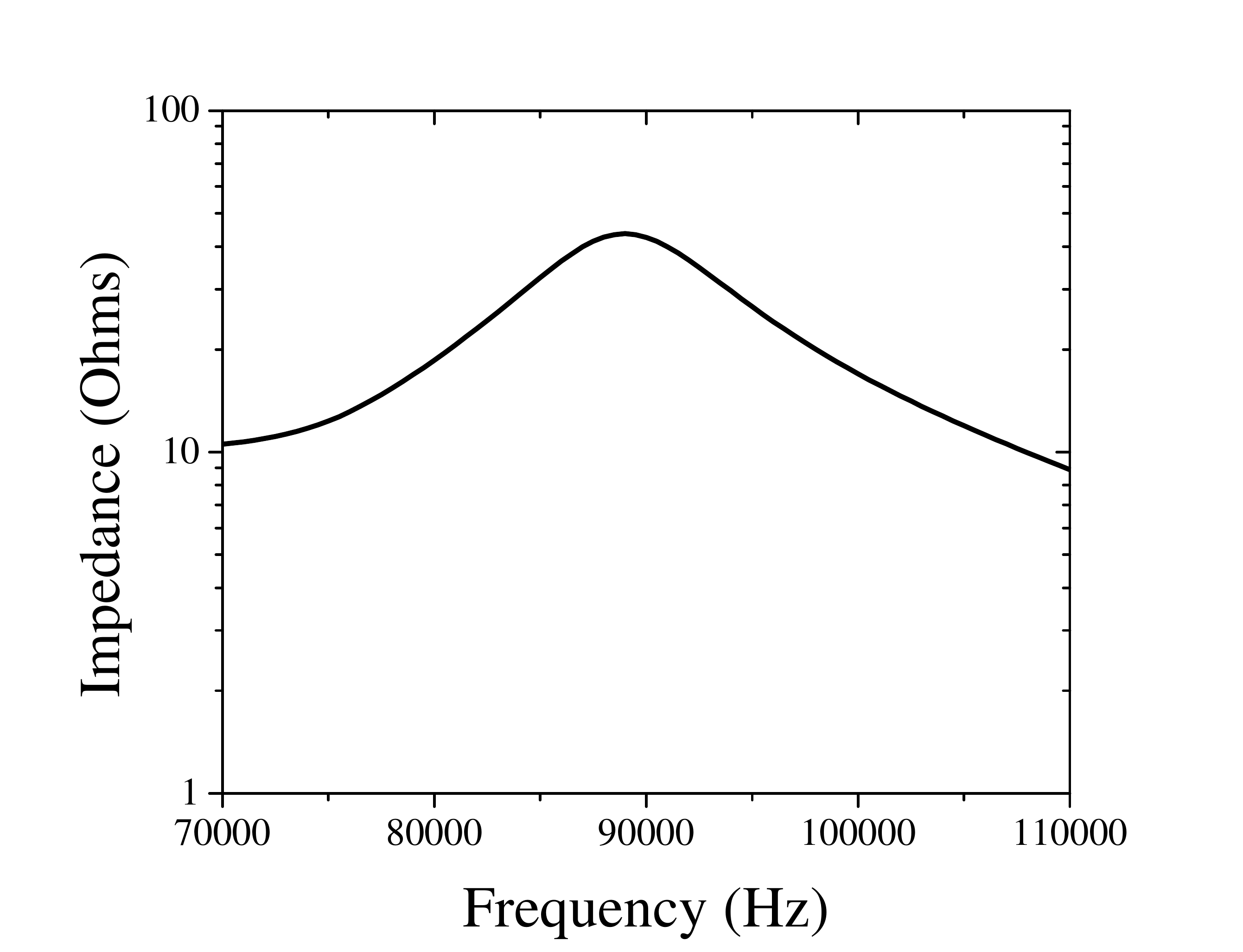} \;
    (d)\includegraphics[width=0.9\columnwidth]{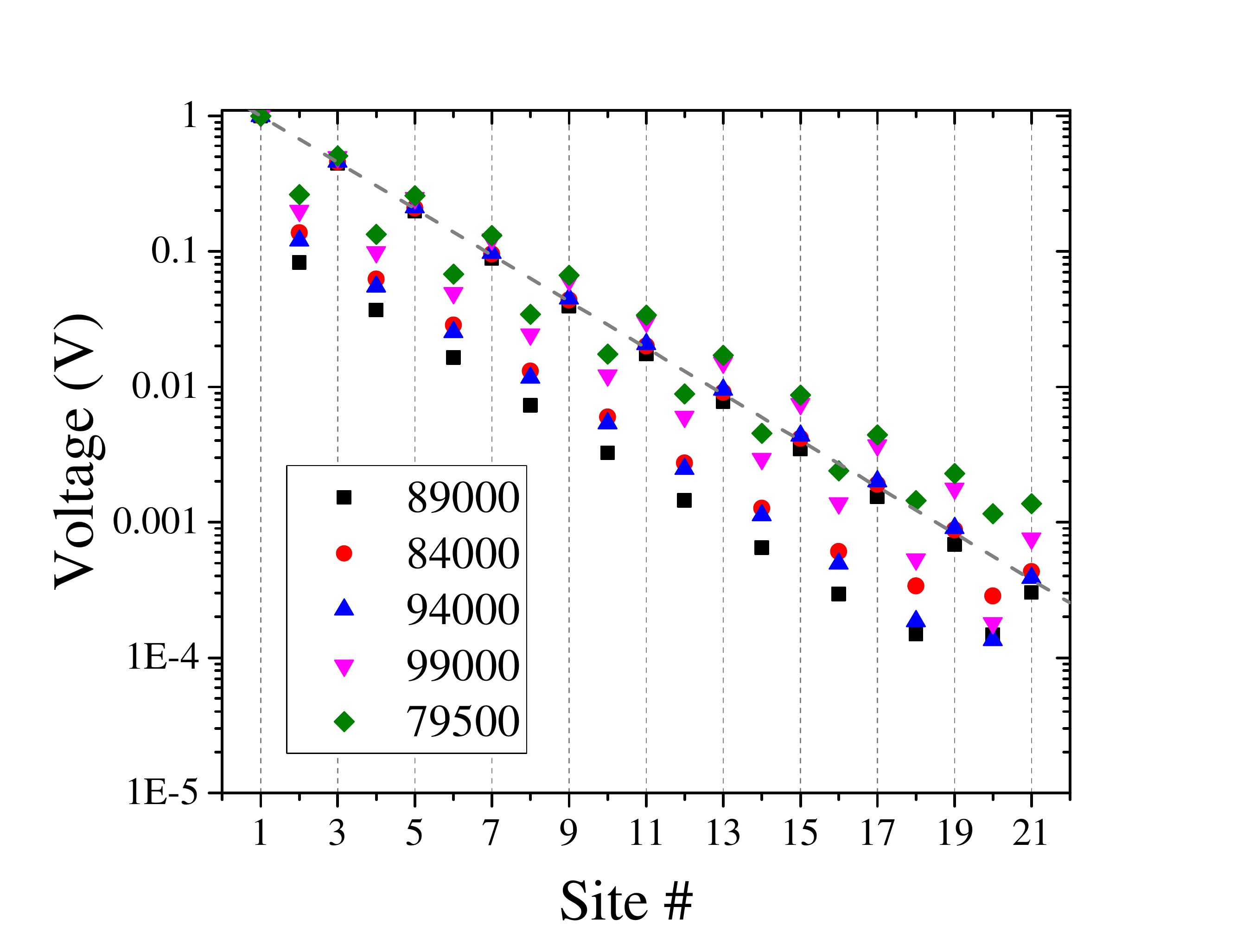} \\
    \caption{Edge state of the SSH memcircuit with in-series connected memristive elements. The edge state can not be clearly distinguished in (a) as its main feature is the suppression of voltages at even sites. However, the edge state can be distinguished in the inverse voltage plot, (b). The parameters used in this simulation are $R_{on}=1$~Ohm, $R_{off}=10$~Ohm, $V_t=0.1$~V, $R(t=0)=1$~Ohm. (c) Impedance as a function of frequency. (d) Voltage distributions at various frequencies. \label{fig:223}}
\end{figure*}

The admittance matrix has then the same structure as that with parallel memristive elements. The same arguments then hold regarding the 
custodial status of the chiral symmetry. The only major difference is that its detection can only be observed at relatively small $R_M$, namely so long as the current in the circuit is not too small to be detected. In this case, the mid-gap state is weakly delocalized compared to the case of in-parallel memristive elements, as shown in 
Fig.~\ref{fig:223}. 


\end{document}